\documentclass{article} % For LaTeX2e
\usepackage{iclr2027_conference,times}
\usepackage{amsmath,amsfonts,bm}

\def\eqref#1{equation~\ref{#1}}
\def\1{\bm{1}}

\DeclareMathAlphabet{\mathsfit}{\encodingdefault}{\sfdefault}{m}{sl}
\SetMathAlphabet{\mathsfit}{bold}{\encodingdefault}{\sfdefault}{bx}{n}

\usepackage{hyperref}
\usepackage{url}
\usepackage{graphicx}
\usepackage{fvextra}
\usepackage{tcolorbox}
\tcbuselibrary{breakable}
\usepackage{tabularx}
\usepackage{xcolor}
\usepackage{booktabs}
\usepackage{enumitem}
\usepackage{colortbl}
\usepackage{capt-of}
\usepackage{pifont}
\usepackage{float}
\usepackage{multirow}

\definecolor{gainblue}{HTML}{1F5FA8}
\definecolor{lossorange}{HTML}{C2410C}
\definecolor{headshade}{HTML}{E8EEF7}
\definecolor{promptchange}{HTML}{FFF1B8}
\newcommand{\up}[1]{\textcolor{gainblue}{$+$#1}}
\newcommand{\dn}[1]{\textcolor{lossorange}{$-$#1}}

\setlist[itemize]{topsep=2pt, itemsep=1pt, parsep=0pt}

\title{Correct Code, Broken Contributions? SWE-CC: Benchmarking Repository Policy Compliance for Coding Agents}

\author{
Truong Hai Dang$^{1}$ \quad
Rayner Goh$^{1}$ \quad
Thanh Le-Cong$^{2}$ \quad
Yintong Huo$^{1}$\thanks{Corresponding author: \href{mailto:ythuo@smu.edu.sg}{\texttt{ythuo@smu.edu.sg}}.} \\
$^{1}$Singapore Management University \qquad
$^{2}$Singapore University of Technology and Design
}

\iclrfinalcopy
\begin{document}

\maketitle
\lhead{}

\begin{abstract}
Autonomous coding agents now resolve a substantial share of real-world GitHub issues. However, passing functional tests differs fundamentally from producing a high-quality contribution acceptable for merging. 
Mature open-source projects publish repository-specific contribution policies, spanning style, git, testing workflows, to ensure code quality and long-term maintainability. Because existing benchmarks evaluate patches solely on unit tests, \textit{agent compliance with repository governance remains unknown.} 
In this paper, we introduce \textsc{SWE-CC}\footnote{\raggedright Benchmark and source code: \href{https://github.com/dangtruong01/swe-cc-arxiv}{\nolinkurl{github.com/dangtruong01/swe-cc-arxiv}}.\par}, a benchmark evaluating code and process compliance in autonomous software engineering. We develop a semi-automated pipeline that converts developer documentation across 12 open-source repositories into 823 machine-checkable atomic policies. SWE-CC introduces two features: 1) lightweight, deterministic checker functions that represent each policy, 2) a comprehensive auditing mechanism that inspects both agent runtime behaviors and final deliverables.
We evaluate the compliance of agent workflows in 500 end-to-end software contribution tasks extended from SWE-bench Verified.
Our evaluation of four LLMs under two agent scaffolds shows that modern agents suffer from coding compliance issues: although agents produce functionally correct patches, they still violate 43.1\% of applicable project policies, with nearly half of all violations occurring during intermediate execution steps. 
These results show that functional correctness does not guarantee real-world readiness, highlighting that future software engineering agents must reliably conform to repository governance to enable safe and trustworthy deployment.
\end{abstract}

\section{Introduction}

Autonomous coding agents have demonstrated strong capabilities in solving diverse real-world software engineering tasks, including code generation~\citep{zhou2026featurebench}, issue resolution~\citep{jimenez2024swe, yang2026swe}, and code translation~\citep{wang2025repotransbench}. This rapid progress shows agents' ability to produce functionally correct code, measured by passing pre-existing test suites.

While recent studies have begun examining non-functional properties such as execution efficiency~\citep{ma2025swe} and security~\citep{lian2026ase, wang2026realsec}, real-world software engineering equally demands adherence to \textbf{project-level code governance}. In practice, mature software projects define repository-specific contribution policies governing code conventions, commit structures, and documentation standards~\citep{elazhary2019not}. These policies are essential for preserving code quality and long-term maintainability~\citep{gousios2015work, allamanis2014learning}.
Recent reports indicate that non-compliant AI contributions increasingly overwhelm maintainers with review overhead~\citep{li2025rise,stenberg2025slop, newstack2026maintainers}. Beyond that, ungoverned agents pose significant operational threats: 
an agent dropped a production database during an explicit code freeze~\citep{aiid2025replit}. Auditing repository governance is therefore an indispensable prerequisite for safe and reliable agent deployment.

\begin{figure}[t]
\centering
\includegraphics[width=\textwidth]{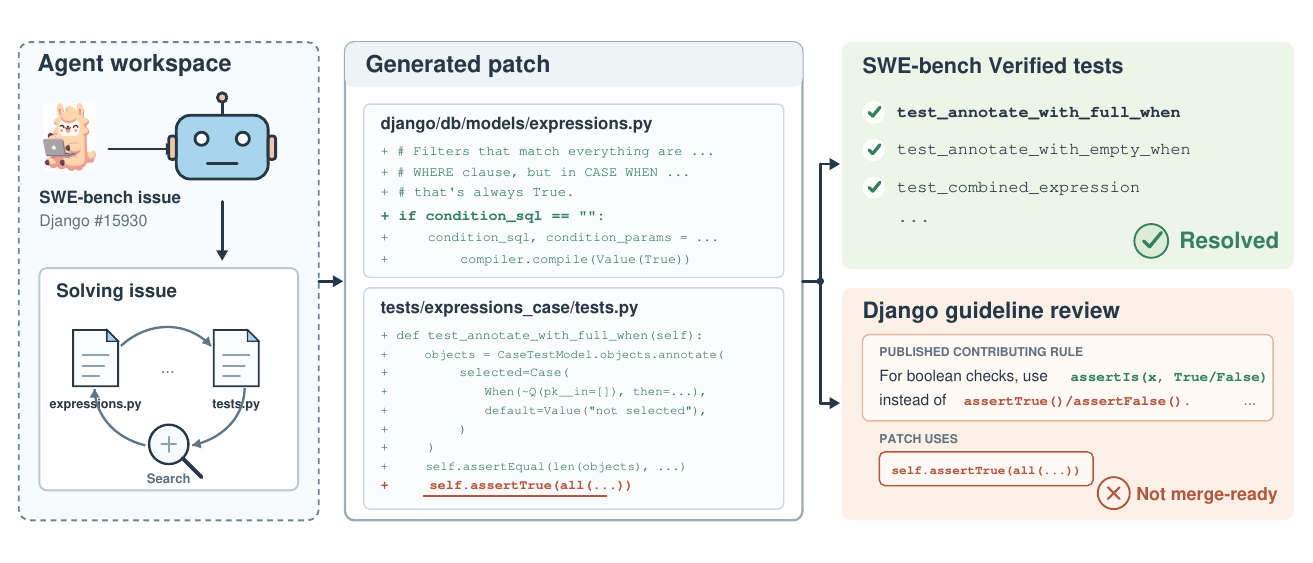}
\vspace{-3mm}
\caption{A SWE-bench-resolved patch violating repository policies. The generated patch passes the full test suite, but violates Django's guideline requiring \texttt{assertIs} for Boolean assertions.}
\label{fig:case}
\vspace{-4mm}
\end{figure}

We demonstrate one project-level policy compliance failure in SWE-bench~\citep{jimenez2024swe}, the widely adopted benchmark for evaluating LLM-based software engineering agents. Figure~\ref{fig:case} illustrates an actual run: the agent successfully navigates the repository and produces a patch that passes the entire test suite, but violates the project's contribution guidelines. While Django's development documentation explicitly mandates \texttt{assertIs(x, [True|False])} for Boolean assertions, the generated patch uses \texttt{self.assertTrue(all(...))}. Because benchmark evaluation solely relies on test execution, such policy violations are not detected, hiding the gap between high benchmark resolve rate and high-quality, mergeable pull requests in practice.

In this paper, we propose a semi-automated pipeline that converts project documentation into a set of machine-checkable policies, and instantiate it with
SWE-CC (\underline{C}ode\underline{C}ompliance), the first benchmark for systematically evaluating \textit{repository-level policy compliance in coding agent behaviors}. 
In practice, an effective agent must not only write functional code, but also proactively discover, interpret, and conform to these governance rules throughout its workflow.
\textsc{SWE-CC} contains 823 real policies written in contributing guidelines and developer documentation across 12 representative repositories, covering main software development lifecycle.

\textsc{SWE-CC} has two core design principles. First, it operationalizes natural-language policies into \textit{lightweight checker functions}. This executable representation converts guidelines that require labor-intensive manual review into objective and reproducible evaluation criteria. It also ensures that repository-level policies are applied across all tasks within that repository, enabling both scalable evaluation and fine-grained violation diagnostics. 
Second, it audits both \textit{runtime behavior} and \textit{final deliverables}. Rather than inspecting only the submitted patch, \textsc{SWE-CC} records the agent's full execution trajectory alongside its terminal outputs, executing checker functions directly over these artifacts to capture policy violations that occur throughout the workflow.

%%% results and findings
Using SWE-CC, we systematically evaluate leading coding agents across diverse foundation models and scaffolding frameworks on 500 end-to-end software contribution tasks modified from SWE-bench Verified. Our empirical analysis reveals several key findings. 
\ding{202} \textbf{Policy violations are pervasive across current agents.} Even the most recent frontier models (e.g, GPT-5.6 Luna, released in July 2026) violate nearly half ($44.7\%$) of the repository policies, exposing substantial deployment risks in organizational production environments.
\ding{203} \textbf{Half of the compliance issues arise during intermediate execution.} 50.3\% of the policy violations occur prior to final commit packaging (e.g., improper test execution), which can only be detected by auditing agent execution trajectories.
% Among the patches that pass functional test suites, 99.8\% violate at least one documented repository policy; the model ranking second in task resolution falls to last in compliance.
\ding{204} \textbf{Non-compliance stems from both discovery failures and reasoning gaps.} Even when explicitly instructed to search, agents initiate policy searches in only 28.9\% of executions. With processed policies provided in context in advance, agents gain an 8.75\% compliance rate on average.

\paragraph{Contributions.} This work introduces SWE-CC, a practical and comprehensive benchmark for evaluating AI coding compliance against its repository-level policies.

\begin{itemize}[leftmargin=*]
\item \textbf{Pipeline.} We present an AI-assisted benchmark construction pipeline that mines real-world project policies and synthesizes them into deterministic, executable checker functions for agent auditing.
\item \textbf{Benchmark.} We propose SWE-CC, the first project-dependent code compliance benchmark to assess agents' ability in reasoning and conforming to coding policies, spanning 823 distinct policies across 12 repositories. 
\item \textbf{Empirical findings.} We benchmark frontier coding agents under two policy-provision settings and two scaffolds. Experimental results reveal that agents violate 43.1\% of applicable project policies, with nearly half of the violations occurring during execution trajectory, highlighting a critical limitation in building policy-compliant agents.
% \item \textbf{Implications.} We call for community-wide efforts toward policy-compliant coding
% agents by adopting \textsc{SWE-CC} and its pipeline for new repositories, auditing how agents work
% alongside what they submit, and training agents to find and follow project policies, so that coding
% agents are held to the same standards as the human contributors they work beside.
\end{itemize}

\section{Related Work} 
\label{sec:related}
% Code agent benchmarks, conformance/rule-checking work, agent instruction-following.
Most software engineering benchmarks assess an agent's capability to generate functionally correct code by executing a project's existing test suite. 
Following SWE-bench~\citep{jimenez2024swe}, subsequent work has broadened it along several directions, including multiple programming language evaluation, synthesizing training data, multi-bug maintenance~\citep{rashid2025swe, miserendino2025swe, yang2026swe, pan2024training, zhou2026featurebench, tang2026devops}.
% , and larger units of development work than a single issue
More recently, there exist benchmarks evaluating non-functional properties, such as code efficiency~\citep{ma2025swe}, security~\citep{lian2026ase, wang2026realsec}, and general code quality (SWE-NFI~\citep{xue2026swe}). Among them, SWE-Gate~\citep{he2026swe} and SWE-SHIELD~\citep{yu2026does} are the two closest to our work, where they mine implicit review criteria from historical pull request discussions to judge patch acceptance. RepoComplianceBench~\citep{yang2026first} restricts its evaluation scope to AI-assistance disclosures, excluding the broader spectrum of repository governance.
Table~\ref{tab:vs-prior} compares SWE-CC with existing non-functional coding evaluation benchmarks.

However, existing evaluations exhibit two critical limitations: 1) Scope of governance. They evaluate only the final code patch, neglecting the intermediate agentic workflow. An agent may produce functional code yet violate contribution standards, such as PR metadata disclosure, or required testing workflows.  2) Policy provision. Prior studies inject the processed rules into the context window, but real-world autonomous engineering demands \textit{in-situ} policy retrieval process, as governance rules are distributed across the repository. \textsc{SWE-CC} addresses these limitations by auditing agent workflow against a project's full documentation, requiring agents to autonomously navigate and reason over policies in-the-wild.

\begin{table}[t]
\centering
\scriptsize
\setlength{\tabcolsep}{4pt}
\caption{Comparing \textsc{SWE-CC} against the closely-related benchmarks. \emph{Evaluation scope} is the evaluation subject, and \emph{Policy source} is where its policies come from. \emph{Agent behavior} marks whether the agent's trajectory is audited, and \emph{Policy retrieval} indicates whether the agent must find the policies itself ($\circ$ not supported, $\bullet$ supported). \emph{Policy evaluator} records how compliance is evaluated.}
\label{tab:vs-prior}
\begin{tabular}{@{}lllcccc@{}}
\toprule
Benchmark & Evaluation scope & Policy source
 & \# Policy & \shortstack{Agent\\behavior}
 & \shortstack{Policy\\retrieval} & \shortstack{Policy\\evaluator} \\
\midrule
SWE-NFI                        & Patch quality    & Literature, common practice   & 92 & $\circ$ & $\circ$   & Executable \\
SWE-Gate                 &  Patch quality   &  Code review history  & 303 & $\circ$ & $\circ$   & Executable \\
RepoComplianceBench  & AI declaration  & Contribution file  & 455   & $\circ$ & $\bullet$ & Executable + LLM \\
SWE-SHIELD             & Patch quality  & Code review history  & 1,787 & $\circ$ & $\circ$   & LLM-as-a-judge \\
\midrule
\textsc{SWE-CC} (ours)                    & Coding workflow & Project documentation & 823 & $\bullet$ & $\bullet$ & Executable \\
\bottomrule
\end{tabular}
\end{table}

\section{SWE-CC}
\label{sec:swecc}

We present a pipeline that compiles project documentation into a corpus of executable policy checkers, and instantiate it on the twelve repositories of SWE-bench Verified. \S\ref{sec:construction} describes the benchmark construction process, including policy extraction and checker construction. \S\ref{sec:benchmark-implementation} and \S\ref{sec:stats} details implementation and summarizes the benchmark characteristics, respectively. \S\ref{sec:task} and \S\ref{sec:metrics} define the evaluation protocol and its metrics.

\textbf{Problem Formulation.} SWE-CC evaluates whether a coding agent adheres to repository policies across its runtime behavior and final deliverables. Given a task instruction $T$ and a repository environment $E$, an agent yields an execution trajectory of sequential steps $\tau = (t_1, t_2, \dots, t_N)$ and a final task output $C$. 
For each repository, \textsc{SWE-CC} formalizes project governance into a set of deterministic checker functions $\mathcal{F} = \{f_1, f_2, \dots, f_K\}$. Each function $f_k(\tau, C) \in \{0, 1, \bot\}$ evaluates the agent's performance by examining runtime behaviors $\tau$ and final code artifacts $C$. $\bot$ indicates the policy is not applicable (not triggered) in this execution.

\begin{figure}[t]
\centering
\includegraphics[width=\textwidth]{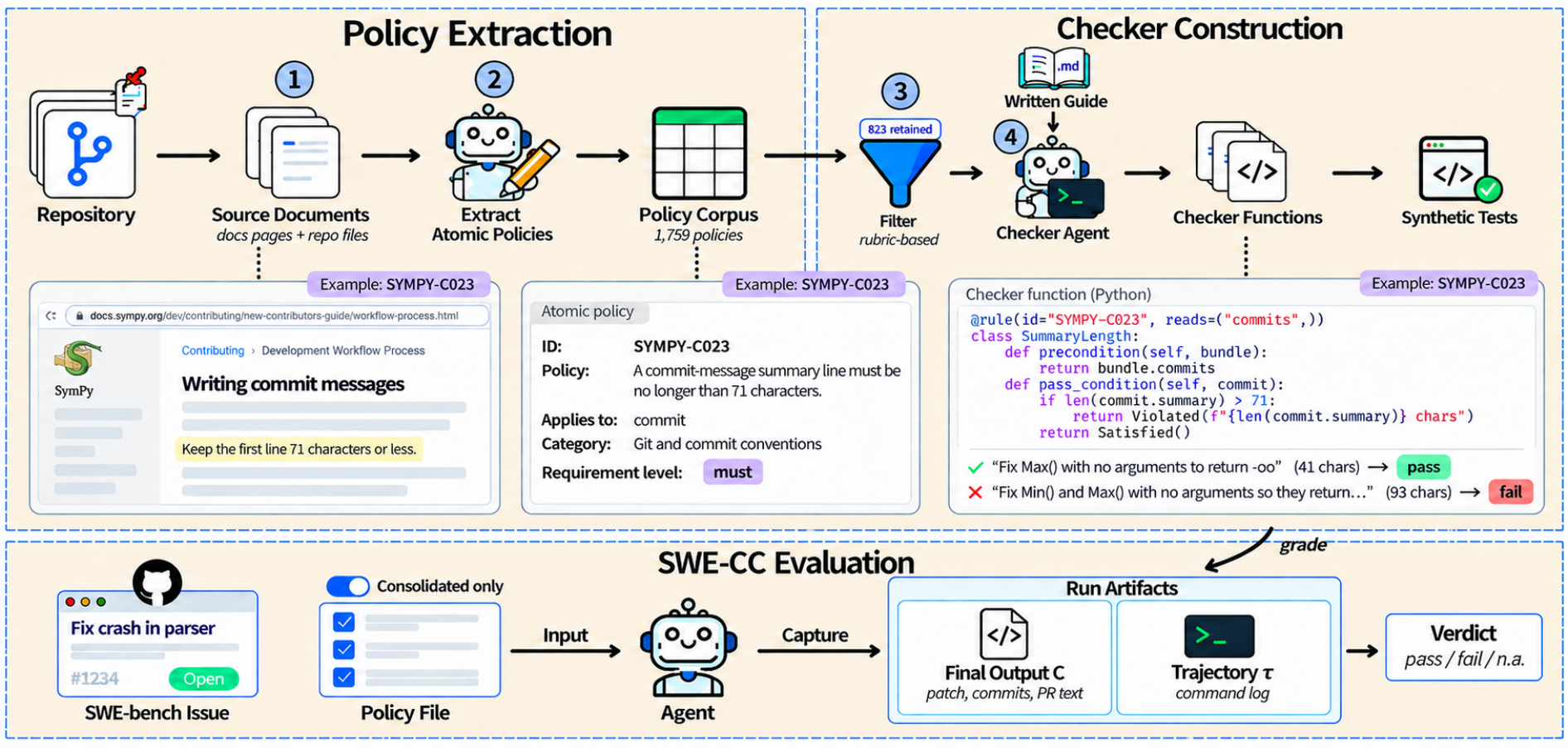}
\caption{\textsc{SWE-CC} benchmark construction pipeline. An LLM agent extracts atomic policies from
project documentation, filters them to 823 checkable, mandatory policies, and compiles each into a
checker function. On a SWE-bench issue, the checker functions grade the agent's final output and
trajectory. We illustrate this with one policy example: commit summary $\le$ 71 characters.}
\label{fig:overview}
\end{figure}

\subsection{Benchmark Construction Pipeline}
\label{sec:construction}

\subsubsection{Policy Extraction}
\label{sec:extraction}

\textbf{Step 1. Sourcing policy documents.} Given an open-source project, the construction starts with identifying the policy documents written for developers and contributors. To do this, we locate the latest published version of the project’s developer documentation, start from its contributor entry page, and traverse the documentation navigation to identify and retain pages intended for contributors. 
% The published documentation, however, does not carry every policy a contributor is held to. 
Since these documents are distributed across different locations, we also collect policy-included files in the repository, such as the pull request template (\texttt{.github/PULL\_REQUEST\_TEMPLATE.md}), which pre-fills the actions of every new pull request, and instruction files addressed to coding agents (\texttt{AGENTS.md}). 
% We collect these documents from the repository manually. 

\textbf{Step 2. Extracting atomic policy statements.} From the identified sources, an LLM agent extracts each contributor obligation as a structured policy specifying a \textit{precondition} (when the rule applies) and a \textit{post-condition} (the required behavior).
As a documentation sentence often compounds several obligations, we ask the agent to decompose them into independent, atomic policies so that each downstream checker maps to exactly one rule. To maintain verifiability and transparency, we also instruct our LLM agent to keep sources of each policy, including its original sentence and location in the documentation, and relevant context (explicit exceptions or designated tools) that separates mandatory requirements from recommendations.
% Appendix~\ref{app:pipeline} gives the full procedure.

% \textbf{Implementation.} Our policy extraction was executed by Claude Opus 5 at high reasoning effort per project, each was run independently, following a written procedure that fixes the traversal, the criterion for retaining a page, and the fields of a policy row (Appendix~\ref{app:prompt-extraction}). The procedure names no project and is issued unchanged to all twelve; a project enters it through two parameters, the documentation root and the repository URL. Every page the agent reads is fetched and recorded in a per-project manifest, and page sources are taken at the commit the documentation build was produced from, so the text a policy is extracted from and the page a contributor reads are identical. An author confirms the page list before extraction begins and the extracted rows before Step 3, after which the rows are frozen, so a policy later disputed is traceable to the page it was read from.

\subsubsection{Checker Construction}
\label{sec:checking}

\textbf{Step 3. Filtering checkable policies.}
\label{sec:classification} We convert each natural-language policy into a deterministic checker function, that is, a predicate evaluated over the agent’s execution trajectory and final code deliverables. To ensure the quality of the policies for verification, we filter them based on the three criteria below.
% However, not all extracted policies are suitable for such automated verification. We therefore use an LLM agent to filter out the policies based on three dimensions below.
% classify each extracted policy along three dimensions and retain only those practical for automated checking.

First, each policy must target an inspectable evidence (e.g., code diff, commit message, or execution trace) and define non-subjective pass/fail conditions.
Second, rules must govern actions directly within an autonomous contributor's purview. We filter out admin-level governance and general end-user instructions.
Third, we restrict the benchmark to mandatory requirements, excluding suggestions or advisory recommendations. Specifically, an LLM assigns labels for these three dimensions with written rubrics.
% (Appendix~\ref{app:rubrics}). 
% This filtering leads to  823 out of the 1{,}759 extracted policies. 

% Excluded policies remain in the released corpus, labelled with the reason for their exclusion.

% the submitted files and the commit message (\textit{output}), the agent's execution trajectory, and non-subjective passing criteria.
% (\textit{trajectory}). A policy such as \emph{write readable code} names no artifact
% (\textit{judgment}) and is excluded. 
% records whether a run produces the artifact a
% policy needs; policies governing code review, continuous integration and branch history concern
% activities our setting does not observe. 

% For each policy, the LLM agent assigns labels for these three dimensions, following written rubrics for determining scope and requirement level (Appendix~\ref{app:rubrics}).
% We keep a
% policy only if it names an artifact, lies in scope and is mandatory, which leaves 823 of the 1{,}759
% extracted policies. 
% Excluded policies remain in the released corpus, labelled with the reason for their exclusion.

\textbf{Step 4. Compiling policies into checker functions.} 
Next, each remaining policy is compiled into one executable and programmatic checker function. This function is represented as a pair of predicates: (i) a \textit{precondition} that identifies the targets the policy applies to, such as each commit the agent made, each file it changed, or each command it ran, and (ii) a \textit{post-condition} judges every target as satisfied or violated, and as undetermined when the artifacts cannot provide sufficient evidence. 
The condition component of an atomic policy determines the precondition, while its required-behavior component determines the post-condition.
For each repository, we synthesize these checkers using an LLM guided by a written procedure that tailors for such translation.

Modern test-driven development practices~\citep{mathews2024test, huang2023agentcoder} have shown code generation improvements through effective test cases and iteration. 
Motivated by this, we validate each checker against three synthetic unit tests: one that satisfies the policy, one that violates it, and one on which the precondition selects nothing. Checkers that fail any test are fed into an iterative refinement. This automated process helps fix initial generation issues and improves checker quality. 
Because the finalized checkers are lightweight and do not invoke an LLM-as-a-judge at evaluation time, they are highly scalable and strictly reproducible.
% No model is invoked at grading time, which makes a compliance figure reproducible from the stored artifacts alone. 
% Appendix~\ref{app:prompt-checkers} describes the detailed process.

% This is necessary because compliance concerns both the resulting artifacts and the process used to produce them, for example, whether the agent ran the necessary test suite before committing.
% To capture this process-level evidence, SWE-CC records an \textit{execution trajectory}: a step-by-step trace of the commands executed by the agent, which are widely used to audit agent behaviour~\citep{wang2026trajaudit,deshpande2025trail}
% Accordingly, SWE-CC evaluates two complementary sources of evidence: 
% (i) the deliverables include the patch, the commits and their messages, and the pull request description; and (ii) the trajectory captures every action the agent took. 

% SWE-CC also distinguishes committed content from uncommitted changes within workspaces, since projects may state policies on what a commit contains.

% The synthetic inputs confirm that a function behaves as written. 
% However, each checker function and its test inputs are generated by the same agent, so passing tests establish only consistency with the agent's interpretation of the policy.
% A model first reviews all 823 functions against a written rubric, which gives full coverage of the corpus and identifies the strata a human sample is drawn from. 

\textbf{Step 5. Validating checker functions.}
While the synthetic tests provide an initial automatic quality review for the checkers, we further validate the fidelity of the checker synthesis pipeline through expert human review.
Two annotators with professional software engineering experience inspected a sample of checkers.
For each checker, annotators were provided with the original policy excerpt, its surrounding repository context, and a formal evaluation guidelines. Each annotator works independently, without sight of the model's verdicts, assigning an \textit{accept} or \textit{reject} judgment along with failure justifications. 

\subsection{Implementation Details and Running Example}
\label{sec:benchmark-implementation}

To instantiate the pipeline, we select 12 representative open-source repositories aligned with SWE-bench Verified~\citep{jimenez2024swe}, constructing the \textsc{SWE-CC} benchmark. From their contributor documentation, the pipeline initially extracts 1,759 atomic policies. Among these, 989 govern inspectable development artifacts or runtime behaviors, and 823 ultimately pass our quality and scope filters to be compiled into executable checker functions. 

\textbf{Implementation.} Each repository is processed by Claude Opus 5~\citep{anthropic2026opus5} at high reasoning effort given the repository and its documentation pages. The two labelling rubrics were drafted and iterated on two projects first, and applied to the rest. 150 out of 823  (18.2\%) checkers are manually examined for validation. The annotators agree on 94.0\% of verdicts (Cohen's $\kappa = 0.72$) and accept 87.2\% of the sampled checkers. This substantial inter-rater agreement~\citep{landis1977measurement} and high accept rate demonstrate that our semi-automated pipeline converts repository governance with high fidelity, establishing a reliable and reproducible pipeline for policy auditing at scale~\citep{ahmed2025can, chowdhury2024swebenchverified}. 
% One author examines the extracted atomic policies.
Appendix~\ref{app:pipeline} details each construction step, and Appendix~\ref{app:rubrics-prompts} gives the labelling rubrics, review guidelines and prompts.

\textbf{Running example.}
The case in Figure~\ref{fig:overview} traces a real policy processing through the pipeline. The SymPy development documentation states:
``Keep the first line 71 characters or less'' (Step \ding{172}). The policy extraction identifies this statement as
an atomic policy whose precondition is that the agent creates a commit and whose post-condition is
that the commit summary line contains at most 71 characters (Step \ding{173}). The filter reserves the
policy as it is output-based, in scope, and mandatory (Step \ding{174}). The resulting checker function selects
every commit produced by the agent and measures the length of its summary line (Step \ding{175}). A
41-character summary satisfies the policy, and a 93-character summary violates it.

\subsection{Benchmark Characteristics}
\label{sec:stats}

\begin{figure}[t]
\centering
\begin{minipage}[t]{0.48\textwidth}
\vspace{0pt}
\centering
\small
\captionof{table}{What the 823 mandatory policies govern. Appendix~\ref{app:categories}
gives an example of each category.}
\label{tab:categories}
\begin{tabular}{@{}lrr@{}}
\toprule
Category & \# Policies & (\%) \\
\midrule
Documentation and docstrings    & 224 & 27\% \\
Language and framework style    & 163 & 20\% \\
Tests and test style            & 160 & 19\% \\
Specialized changes             & 107 & 13\% \\
PR and release metadata         &  57 &  7\% \\
Git and commit conventions      &  40 &  5\% \\
AI-assisted contribution policy &  37 &  4\% \\
Code and quality                &  35 &  4\% \\
\midrule
\textit{Total}                           & 823 &      \\
\bottomrule
\end{tabular}
\end{minipage}%
\hfill
\begin{minipage}[t]{0.48\textwidth}
\vspace{0pt}
\centering
\includegraphics[width=\linewidth,height=0.22\textheight,keepaspectratio]{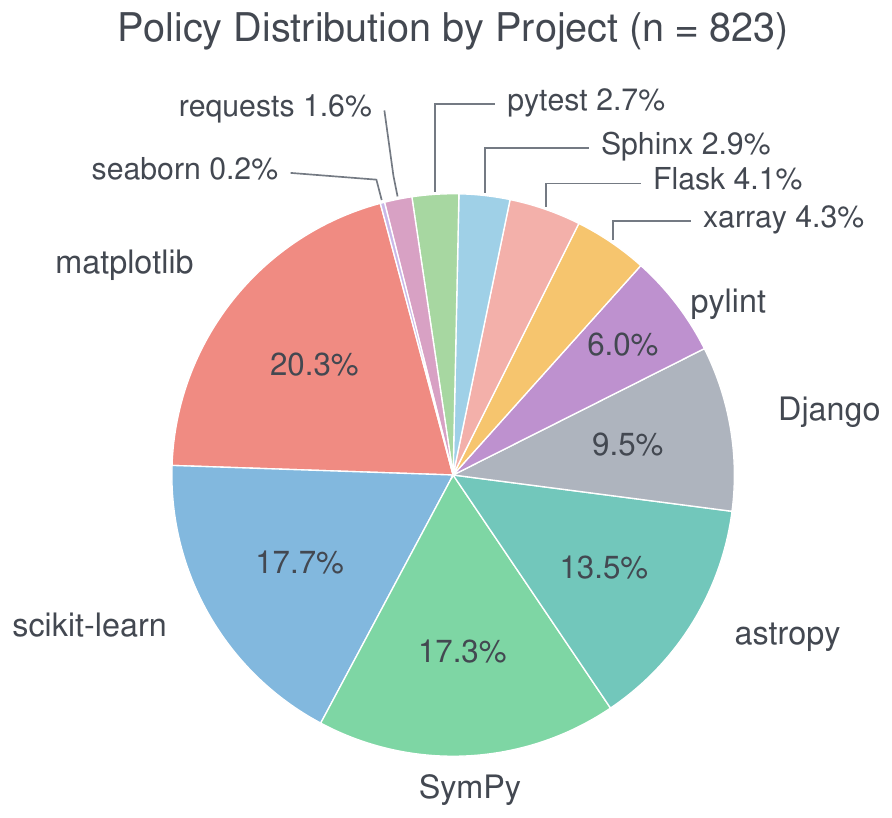}
\captionof{figure}{Distribution of the policies over projects.}
\label{fig:policy-distribution}
\end{minipage}
\vspace{-3mm}
\end{figure}

SWE-CC contains 823 policies over twelve repositories in total, with individual projects contributing between 2 and 167 policies (Figure~\ref{fig:policy-distribution}): Django is 46\% of the instances and 9\% of the policies, matplotlib 7\% and 20\%. Policies also seldom overlap across projects: 12 of the 823 (1.5\%) have similar policies in other projects (Jaccard $\ge 0.6$ over content words). Policies on the same activity can conflict across projects: Django requires an issue number in its commit message, but Flask prohibits one. Repository governance is therefore strictly project-specific. In terms of granularity, each policy articulates a single obligation in 12 words on average, and its checker implementation averages 43 lines of code and 21 lines of unit tests.

To characterize the breadth of policies across the development life cycle, we group the extracted policies into eight governance domains (Table~\ref{tab:categories}). The largest categories govern documentation and docstrings (27\%), language and framework style (20\%) and test style (19\%). The remaining policies cover operational and emerging practices, including git and commit conventions (5\%), AI-assisted contribution policy (4\%) and code quality (4\%). While testing and code quality guidelines appear in all 12 repositories, policies explicitly regulating AI-assisted contributions are already documented in 11 projects, reflecting widespread maintainer concern over unvetted autonomous contributions.

\subsection{Evaluation}
\label{sec:task}

%% tasks
We evaluate agent compliance on 500 \textbf{end-to-end software contribution tasks} extended from SWE-bench Verified~\citep{chowdhury2024swebenchverified}. Unlike existing benchmarks that evaluate an isolated patch~\citep{jimenez2024swe, aleithan2024swe, rashid2025swe, pan2024training}, our setup requires agents to simulate a complete open-source contribution workflow by packaging the solution into a pull request. This encompasses structuring commits, composing informative PR descriptions, and maintaining accompanying tests and documentation alongside the core bug fix.

\textsc{SWE-CC} then audits the agent's contribution workflow against the repository's policies, inspecting both intermediate runtime behaviors and final PR deliverables. The evaluation suite includes two policy provision settings.

%% two settings
\begin{itemize}[leftmargin=*]
\item \textit{Native.} The agent is provided with \textit{the location} of all governing policy documents across the repository (as a URL), and explicitly prompted to look into them before starting. This setting simulates realistic, in-the-wild development where agents must autonomously locate, retrieve, and interpret project guidelines.
\item \textit{Consolidated.}  The \textit{atomic policy statements are aggregated} into a local file (.md) in the environment, with instructions to read it. By removing the search and retrieval bottleneck, this setting isolates the agent's intrinsic ability to reason about and comply with explicit constraints.
\end{itemize}

\begin{table}[t]
\centering
\caption{Triggering, compliance, and resolve rate per model and scaffold. $\Delta$ shows the performance gain of Consolidated. The highest rate per column is in \textbf{bold}, marked separately for Native and Consolidated.}
\label{tab:headline}
\small
\setlength{\tabcolsep}{4pt}
\begin{tabular}{l@{\hspace{1.6em}} ccc @{\hspace{1.4em}} ccc}
\toprule
 & \multicolumn{3}{c}{\textbf{mini-SWE-agent}} & \multicolumn{3}{c}{\textbf{OpenHands}} \\
\cmidrule(lr){2-4}\cmidrule(lr){5-7}
Agent models
  & \textbf{\shortstack{Triggering\\rate (\%)}}
  & \textbf{\shortstack{Compliance\\rate (\%)}}
  & \shortstack{Resolve\\rate (\%)}
  & \textbf{\shortstack{Triggering\\rate (\%)}}
  & \textbf{\shortstack{Compliance\\rate (\%)}}
  & \shortstack{Resolve\\rate (\%)} \\
\midrule
\addlinespace[2pt]
\rowcolor{headshade}
\multicolumn{7}{c}{\textbf{Native Setting}} \\
\midrule
GPT-5.6 Luna   & 23.5 & 51.8 & 78.6 & 24.4 & 55.3 & 86.6 \\
 Gemini 3.7 Flash      & 23.8 & 53.9 & 80.2 & 24.3 & 56.3 & 78.4 \\
DeepSeek V4 Flash    & \textbf{25.6} & \textbf{61.3} & \textbf{93.4} & \textbf{26.0} & \textbf{63.8} & \textbf{94.4} \\
 Kimi K2.5      & 24.6 & 55.9 & 74.6 & 24.8 & 57.2 & 76.2 \\
\midrule
\rowcolor{black!5}
\quad \textit{Average}      & 24.4 & 55.7 & 81.7 & 24.9 & 58.2 & 83.9\\ 

\midrule
\addlinespace[2pt]
\rowcolor{headshade}
\multicolumn{7}{c}{\textbf{Consolidated Setting (with improvement $\Delta$)}} \\
\midrule
GPT-5.6 Luna  & 25.9 \textit{(\up{2.4})} & 67.6 \textit{(\up{15.8})} & 77.6 \textit{(\dn{1.0})} & 26.2 \textit{(\up{1.8})} & 67.5 \textit{(\up{12.2})} & 84.8 \textit{(\dn{1.8})} \\
Gemini 3.7 Flash  & 25.3 \textit{(\up{1.5})} & 65.3 \textit{(\up{11.4})} & 81.2 \textit{(\up{1.0})} & 26.2 \textit{(\up{1.9})} & 66.8 \textit{(\up{10.5})} & 79.4 \textit{(\up{1.0})} \\
DeepSeek V4 Flash  & \textbf{26.3} \textit{(\up{0.7})} & \textbf{67.7} \textit{(\up{6.4})} & \textbf{92.6} \textit{(\dn{0.8})} & \textbf{27.6} \textit{(\up{1.6})} & \textbf{71.2} \textit{(\up{7.4})} & \textbf{93.8} \textit{(\dn{0.6})} \\
Kimi K2.5  & 25.0 \textit{(\up{0.4})} & 56.9 \textit{(\up{1.0})} & 74.0 \textit{(\dn{0.6})} & 26.2 \textit{(\up{1.4})} & 62.4 \textit{(\up{5.2})} & 74.2 \textit{(\dn{2.0})} \\
\midrule
\rowcolor{black!5}
\quad \textit{Average} & 25.6 \textit{(\up{1.2})} & 64.4 \textit{(\up{8.7})} & 81.4 \textit{(\dn{0.3})} & 26.6 \textit{(\up{1.7})} & 67.0 \textit{(\up{8.8})} & 83.1 \textit{(\dn{0.8})}\\
\bottomrule
\end{tabular}
\vspace{-3mm}
\end{table}

\subsection{Metrics}
\label{sec:metrics}
%% parse
Most policies govern specific activities. For example, guidelines for docstring formatting only apply when an agent modifies a function signature. To reflect this, each policy is associated with an applicability precondition, indicating whether the agent's trajectory or deliverable is in scope. We therefore use two complementary compliance metrics alongside standard functional correctness:
\begin{itemize}[leftmargin=*]
\item \textit{Triggering Rate.} The proportion of a repository's policies whose preconditions are activated by the agent's workflow, reflecting how much of the governance rule set is brought into scope.
\item \textit{Compliance Rate.} The proportion of those triggered policies that the agent successfully passes, indicating their capability to conform to policies.
\item \textit{Resolve Rate.} The standard functional success rate from SWE-bench, measuring whether the code patch alone passes the repository's unit tests.
\end{itemize}

\section{Experiments}
%%% how should we organize the evaluation part..
  \subsection{Experimental Setup}
  \label{sec:setup}

We evaluate two representative agent frameworks (i.e., mini-SWE-agent~\citep{yang2024swe} and OpenHands~\citep{wang2026openhands}), each paired with four LLM backbones from different model families and capability tiers: GPT-5.6 Luna~\citep{openai2026gpt56}, Gemini 3.7 Flash~\citep{googledeepmind2026gemini37flash}, DeepSeek V4 Flash 0731~\citep{xu2026deepseek}, and Kimi K2.5~\citep{team2026kimi}. This leads to a total of 8,000 (8*500*2 settings) coding workflows to audit compliance. 

Our experiments follow the default implementations and configurations. We keep the scaffold's original system prompt and tools, and extend only the task prompt, which adds the commit and pull-request steps of \S\ref{sec:task}, permits edits to all files, and takes the pull request as text (Appendix~\ref{app:agent-prompts}). 

Resolution is scored by the SWE-bench suite after extracting the patch contents from model final outputs. SWE-CC evaluates compliance after each execution in 120 to 215 ms on a local MacBook Pro (Apple M4 Pro, 24\,GB unified memory), showing the scalability of our approach.
 
  \subsection{Results}
In this section, we report results for agents with different scaffolds and backbones, then provide analysis and insight into the performance and bottlenecks. To validate the prompt and scaffold implementation, we compare our \textit{Native} resolution rate under mini-SWE-agent against those with the same models and scaffold on SWE-bench Verified~\citep{vals2026swebench}. All experiments use the same infrastructure.

\begin{figure}[t]
\centering
\begin{minipage}[t]{0.71\textwidth}
\vspace{0pt}
\centering
\includegraphics[width=\linewidth]{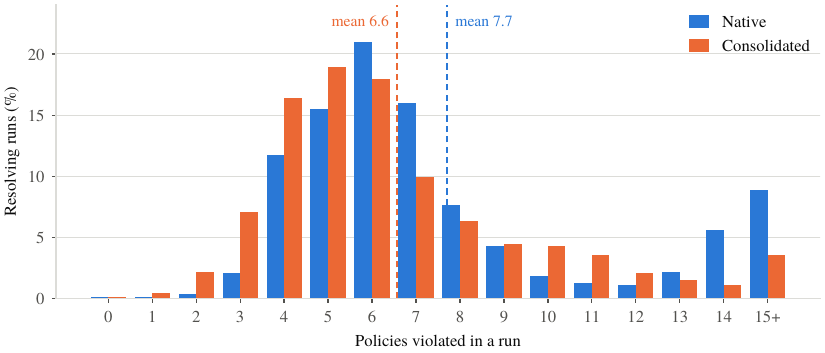}
\captionof{figure}{Policies violated over the functionally-resolved cases. Supplying the policies shifts the distribution left by about one violation, from a mean of
7.7 to 6.6. An execution triggers 22.6 policies on average.} 
%and leaves the zero-violation bar close to empty: 6 of 3{,}312 \textit{Native} runs and 4 of 3{,}288
%\textit{Consolidated} runs violate nothing.}
\label{fig:violations}
\vspace{-3mm}
\end{minipage}%
\hfill
\begin{minipage}[t]{0.25\textwidth}
\vspace{0pt}
\centering
\includegraphics[width=0.95\linewidth]{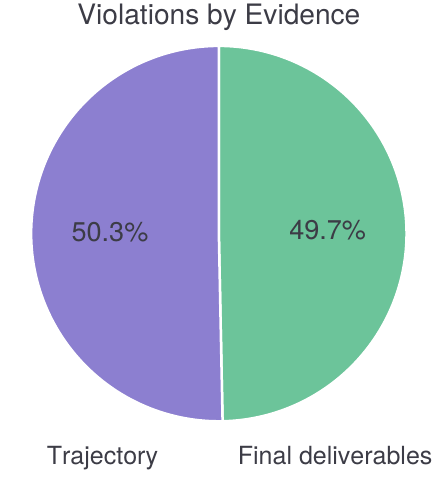}
\captionof{figure}{Whether the violations happen in the final
deliverables or trajectory.}
\label{fig:violations-evidence}
\end{minipage}
\vspace{-1mm}
\end{figure}

\textbf{Modern agents suffer from coding compliance issues.} Table~\ref{tab:headline} summarizes agent performance across both policy provision settings.
Across the board, models exhibit severe compliance deficits. Even the top-performing model achieves a compliance rate of only 63.8\% under the \textit{native} setting and 71.2\% under the \textit{consolidated} setting. This reveals that nearly 30\% of repository policies are violated during development, regardless of how functionally effective their code artifacts are.
Most frontier agents demonstrate strong functional problem-solving ability, generating effective patches in 74.6\%--94.4\% of runs. 
However, agents adhere to only 51.8\%--63.8\% of applicable policies during these runs.  A further examination in Figure~\ref{fig:violations} reveals that agents violate a mean of 7.7 out of 22.6 triggered policies (34.1\%) per execution. With the help of consolidated policies, the mean number of violations drops to 6.6.

\textbf{A large share of these violations happen in the agent's execution}. The trajectory holds 50.3\% of the violations among resolved runs (Figure~\ref{fig:violations-evidence}), such as running the test suite before committing, which can only be checked in agent's trajectory. An evaluation that inspects the final patch alone misses these violations, and \textsc{SWE-CC} captures them by checking the trajectory. These results demonstrate that functional evaluation masks substantial governance failures and underscores the necessity for agent auditing. 

% ===== Table 3: by rule category =====
\begin{table}[t]
\centering
\caption{Triggering and compliance rate by governance domain, averaging over four models within each scaffold. For each domain, the highest compliance rate is in \textbf{bold} and the second highest is \underline{underlined}.}
\label{tab:bycat}
\scriptsize
\setlength{\tabcolsep}{5pt}
\begin{tabular}{llcccc}
\toprule
 & & \multicolumn{2}{c}{\textbf{mini-SWE-agent}} & \multicolumn{2}{c}{\textbf{OpenHands}} \\
\cmidrule(lr){3-4}\cmidrule(lr){5-6}
Category & & \shortstack{Native (\%)} & \shortstack{Consolidated (\%)} & \shortstack{Native (\%)} & \shortstack{Consolidated (\%)} \\
\midrule
Git and commit conventions & Triggering rate & 52.3 & 59.7 & 55.7 & 60.4 \\
 & Compliance rate & 59.9 & \underline{87.0} & 63.4 & \textbf{90.1} \\
\addlinespace
PR and release metadata & Triggering rate & 36.3 & 45.6 & 37.5 & 51.8 \\
 & Compliance rate & 36.4 & \underline{51.6} & 39.6 & \textbf{58.9} \\
\addlinespace
Code and quality & Triggering rate & 83.7 & 83.8 & 82.5 & 82.8 \\
 & Compliance rate & 41.1 & \textbf{56.1} & 41.5 & \underline{54.4} \\
\addlinespace
AI-assisted contribution policy & Triggering rate & 91.6 & 92.8 & 93.2 & 95.9 \\
 & Compliance rate & 15.3 & \underline{26.6} & 20.6 & \textbf{30.7} \\
\addlinespace
Tests and test style & Triggering rate & 26.5 & 26.8 & 26.4 & 26.9 \\
 & Compliance rate & 67.4 & \underline{71.4} & 71.0 & \textbf{74.6} \\
\addlinespace
Language and framework style & Triggering rate & 24.9 & 25.0 & 25.4 & 25.5 \\
 & Compliance rate & 84.0 & \underline{85.1} & 84.2 & \textbf{85.9} \\
\addlinespace
Specialized changes & Triggering rate & 1.2 & 1.2 & 1.2 & 1.2 \\
 & Compliance rate & 70.9 & \textbf{74.2} & \underline{72.8} & 69.3 \\
\addlinespace
Documentation and docstrings & Triggering rate & 5.4 & 6.3 & 6.0 & 6.8 \\
 & Compliance rate & \textbf{62.7} & 61.1 & 62.2 & \underline{62.6} \\
\bottomrule
\end{tabular}
\vspace{-3mm}
\end{table}

\textbf{Feeding the processed rules boosts compliance performance, but the improvement is not enough.}
Comparing consolidated with native performance reveals a consistent increase in compliance across all models, with an average gain of 8.75\%. 
The greatest improvements are under GPT-5.6 Luna, where mini-SWE-agent gains +15.8\% and OpenHands gains +12.2\%. 
Meanwhile, functional resolution rates fluctuate within 2\%, showing that policy provision does not degrade problem-solving capability. 
Nevertheless, compliance remains a limitation: even under the best configuration, the top model violates about 30\% policy, highlighting the demands for a better compliance-enhanced strategy.

\textbf{Compliance failures over governance domains.}
Table~\ref{tab:bycat} breaks down triggering and compliance rates across the eight policy categories.
(1) Policies governing AI-assisted contributions (92\%) and code \& quality (82\%) trigger frequently, as both apply broadly to repository-level code edits. Conversely, documentation and docstrings and specialized changes trigger rarely, as bug-fixing benchmarks seldom require modifying project-wide documentation.
(2) Compliance rates differ by type. In OpenHands, policies governing language and framework style see the highest native compliance, which changes minimally when guidelines are provided (+1.7\%). Because pre-trained models naturally adopt idiomatic code formatting, they satisfy these intrinsic conventions by default.
(3) Ironically, despite being specifically authored to govern AI behavior, \textit{AI-assisted contribution policies} exhibit the lowest compliance (15\% in native). While providing consolidated guidelines, agents still fail to include mandated AI disclosures, demonstrating that agents often omit mandated self-disclosures.

% ===== Table : did the rules arrive =====
\begin{table}[t]
\centering
\caption{The portion of the policies that reach model context.
\textit{Native} columns are progressive: the agent \emph{attempted} to fetch a policy document
and \emph{retrieved} policy text from it (highest in \textbf{bold}, second \underline{underlined}).
\textit{Consolidated} reports whether the mounted file was \emph{opened} and
what share of its content was \emph{delivered} into context.}
\label{tab:arrival}
\scriptsize
\setlength{\tabcolsep}{5pt}
\begin{tabular}{ll cc cc}
\toprule
 & & \multicolumn{2}{c}{\textbf{Native} (autonomously discover)} & \multicolumn{2}{c}{\textbf{Consolidated}} \\
\cmidrule(lr){3-4}\cmidrule(lr){5-6}
Model & Scaffold & Attempt (\%) & Retrieve (\%)  & Opened (\%)  & Delivered (\%)  \\
\midrule
GPT-5.6 Luna & mini-SWE-agent & 50.0 & 13.2 & 100.0 & 100.0 \\
 & OpenHands & \textbf{75.6} & \textbf{43.2} & 100.0 & 99.9 \\
\addlinespace
Gemini 3.7 Flash & mini-SWE-agent & 10.0 & 0.4 & 100.0 & 100.0 \\
 & OpenHands & 2.8 & 1.0 & 98.6 & 100.0 \\
\addlinespace
DeepSeek V4 Flash & mini-SWE-agent & 34.8 & 10.6 & 100.0 & 99.7 \\
 & OpenHands & \underline{51.0} & \underline{31.6} & 100.0 & 100.0 \\
\addlinespace
Kimi K2.5 & mini-SWE-agent & 3.6 & 0.2 & 42.6 & 97.9 \\
 & OpenHands & 3.6 & 0.6 & 91.4 & 99.6 \\
\midrule
\rowcolor{black!5}
\multicolumn{2}{l}{\textit{Average}} & 28.9 & 12.6 & 91.6 & 99.6 \\
\bottomrule
\end{tabular}
\vspace{-3mm}
\end{table}

\textbf{Bottlenecks exist in the policy retrieval and reasoning process.} \textit{Native} runs require the agent to discover and fetch contribution guidelines by itself, where we summarize the performance bottleneck in two causes (Table~\ref{tab:arrival}).
First, 71.1\% of executions do not attempt to even fetch the document on average. 
Second, more than half of the attempts retrieve no policy text, through the harness output cap and the agent's failures in handling the response, such as encoding errors on non-ASCII output. 
Attempt rates also vary significantly across models. 
The \textit{Consolidated} setting eliminates the retrieval bottleneck by mounting all atomic policies directly into a local skill-style file. On average, 91.6\% of executions open the file, bringing 99.6\% of its content into the context window. 
Nevertheless, compliance remains low despite full in-context visibility, demonstrating that governance failures stem from both retrieval and reasoning processes.

\section{Conclusion}
We present a pipeline that converts a repository with established contributor guidelines into a set of executable checker functions, and build \textsc{SWE-CC}, a benchmark that evaluates agentic coding systems on policy compliance across both execution trajectories and final patch outputs. The evaluation revealed a critical
limitation of current coding agents: models violate a large fraction of project guidelines when making software contributions, with roughly half of these violations occurring during intermediate execution.
Further analysis shows that compliance failures stem from bottlenecks in both policy retrieval and reasoning. 

% and the scaffold determines how often the rules are reached while compliance follows the model. 
These results underscore the importance of evaluating the procedural compliance of agent actions, beyond the correctness of their final code artifacts. Moving forward, we will expand \textsc{SWE-CC} across diverse model architectures, scaffolds, and repository benchmarks to jointly measure correctness and compliance.
Ultimately, reporting compliance alongside functional correctness paves the way toward safe, deployable agentic software engineers that align with real-world developer norms and reduce human maintenance overhead.

% real contribution can be made to our software engineering community.

\section*{Reproducibility statement}
We release the data and source code at \url{https://github.com/dangtruong01/swe-cc-arxiv}. Full implementation details, configurations, and prompt templates are in the Appendix, to ensure replicability. 

\bibliography{iclr2027_conference}
\bibliographystyle{iclr2027_conference}

\newpage
\appendix
\section{Benchmark Statistics}
\label{app:stats}

This appendix supports Section~\ref{sec:stats}. It gives the corpus of each project and one example policy per category.

\subsection{Per-project Corpora}
\label{app:corpora}

Table~\ref{tab:corpora} gives, for each project, the number of SWE-bench Verified instances, the policies kept at each filtering step, the evidence each checker reads, and the share of approximate checkers (Appendix~\ref{app:exact}).

\begin{table}[H]
\centering
\small
\caption{The twelve corpora of \textsc{SWE-CC}. \textit{Extracted} counts atomic rules from the pages that bind contributors, \textit{in scope} those a run produces evidence for, and \textit{binding} those stating a requirement; the evidence-type columns partition the binding rules. \textit{Approx.} is the share of checks that approximate their rule (Appendix~\ref{app:exact}).}
\label{tab:corpora}
\begin{tabular}{lrrrrrrrr}
\toprule
 & & \multicolumn{3}{c}{rules} & \multicolumn{3}{c}{evidence type} & \\
\cmidrule(lr){3-5}\cmidrule(lr){6-8}
project & instances & extracted & in scope & binding & output & diff. & traj. & approx. \\
\midrule
astropy      &  22 &  255 & 129 & 111 & 103 &  5 &  3 & 85\% \\
django       & 231 &  143 &  89 &  78 &  67 &  8 &  3 & 55\% \\
matplotlib   &  34 &  293 & 199 & 167 & 148 & 14 &  5 & 82\% \\
scikit-learn &  32 &  274 & 169 & 146 & 131 &  8 &  7 & 83\% \\
sympy        &  75 &  279 & 188 & 142 & 112 & 11 & 19 & 24\% \\
pylint       &  10 &  114 &  59 &  49 &  40 &  1 &  8 & 47\% \\
xarray       &  22 &  139 &  47 &  35 &  27 &  4 &  4 & 51\% \\
flask        &   1 &   88 &  36 &  34 &  20 & 10 &  4 & 65\% \\
sphinx       &  44 &   62 &  27 &  24 &  15 &  2 &  7 & 58\% \\
pytest       &  19 &   66 &  27 &  22 &  14 &  2 &  6 & 59\% \\
requests     &   8 &   41 &  15 &  13 &   6 &  1 &  6 & 77\% \\
seaborn      &   2 &    5 &   4 &   2 &   1 &  0 &  1 &  0\% \\
\midrule
total        & 500 & 1{,}759 & 989 & 823 & 684 & 66 & 73 & 64\% \\
\bottomrule
\end{tabular}
\end{table}

\subsection{Policy Categories}
\label{app:categories}

Table~\ref{tab:category-examples} gives one policy per category, quoted from the released corpus with its
identifier. The eight categories are a closed list fixed before extraction, so that a category means the same
thing in every project.

\begin{table}[h]
\centering
\caption{One mandatory policy per category, as extracted. Counts are repeated from Table~\ref{tab:categories}.}
\label{tab:category-examples}
\small
\begin{tabular}{p{3.1cm}rp{8.2cm}}
\toprule
category & \# & example policy \\
\midrule
Documentation and docstrings & 224 & \textit{Give every public class, method, and function a docstring.} (ASTROPY-C085) \\
Language and framework style & 163 & \textit{Do not raise the bare \texttt{Exception} class.} (ASTROPY-C100) \\
Tests and test style & 160 & \textit{Name test modules \texttt{test\_*.py} or \texttt{*\_test.py}.} (ASTROPY-C002) \\
Specialized changes & 107 & \textit{Keep in-repository data files under about 100\,kB and host anything larger off the repository.} (ASTROPY-C092) \\
PR and release metadata & 57 & \textit{Add a changelog fragment under \texttt{docs/changes/<sub-package>/} describing your change.} (ASTROPY-C061) \\
Git and commit conventions & 40 & \textit{Phrase commit subject lines in past tense and end them with a period.} (DJANGO-C046) \\
AI-assisted contribution policy & 37 & \textit{Disclose any AI tools used in preparing a contribution and what each was used for.} (DJANGO-C057) \\
Code and quality & 35 & \textit{Ensure code passes black, blacken-docs, flake8, isort, and zizmor checks cleanly.} (DJANGO-C069) \\
\bottomrule
\end{tabular}
\end{table}

\section{Construction Pipeline Details}
\label{app:pipeline}

This appendix follows the five steps of Section~\ref{sec:construction}. Steps~1 to~3 run as three stages of one agent workflow: a source manifest, extraction proper, and classification. A deterministic setup step comes first, and each stage produces an artifact that the next consumes. The extraction output is frozen before classification begins. Step~4 compiles the classified policies into checkers and Step~5 validates them. The last subsection gives the acceptance checks and the failure modes we observed. The rubrics and prompts used in each step are in Appendix~\ref{app:rubrics-prompts}.

Each repository is processed by a language-model agent working from a fixed written procedure. A short launcher prompt (Appendix~\ref{app:prompt-launcher}) assigns the agent one repository and directs it to the workflow document, whose stage instructions are reproduced in Appendix~\ref{app:prompts}, together with the two classification rubrics (Appendix~\ref{app:rubrics}) and a format-standard workbook. Agents for different repositories run independently, each confined to its own output directory. We chose this design because the alternative does not scale. Contribution documentation for a large project runs to dozens of pages, the obligations in it are diffuse, and a study spanning twelve repositories cannot be assembled by reading them all by hand. The agent does the reading and the drafting, and a human decides what is admitted.

\textbf{Models and execution.} Extraction and classification are run by Claude Opus~5 at high reasoning effort in Claude Code, one independent agent per project, with web retrieval and shell access, in August 2026. The documentation version read for each repository is recorded in its released run log. The rubrics were written against Django and SymPy and applied unchanged to the other ten repositories.

\textbf{Human review.} One of the authors performs the first two checkpoints (Table~\ref{tab:checkpoints}), confirming the page list before extraction starts and the extracted rows before classification begins. The checks at the third checkpoint are automated by an acceptance script. Every run also ends with a short report from the agent giving the audit results, label counts, and anything in the procedure it found wrong, ambiguous, or missing, which the reviewer reads alongside the corpus before the repository is accepted.

\subsection{Setup}

\textbf{Documentation version.} We read each project's developer documentation, which is the version it directs contributors to and carries material describing changes not yet in any release. Each project is pinned at its latest published build, and all sources for a repository are read at that version. A development alias routinely serves pages built from different commits within one run, so the agent halts only on a different release version. Where builds differ, it pins the affected sources to one commit and records that no extracted quotation is affected. Where a project publishes only a moving alias, the version is resolved by hand and the choice recorded.

\textbf{Off-navigation sources.} Several files that carry real obligations are reachable from no page in the documentation, including pull request and issue templates, pre-commit configuration, changelog fragment documentation, and files addressed to coding agents such as \texttt{CLAUDE.md}, \texttt{AGENTS.md} and \texttt{.github/copilot-instructions.md}. Four of the twelve projects publish such a file, and three of them yield policies. Django's file instructs GitHub Copilot as a code reviewer, so none of it applies to a contributor. A deterministic script matches filename patterns against the repository tree through the GitHub API, retrieves every match as raw text rather than through a code host's rendered view, and bundles them into a single file supplied to the agent. The distinction matters because rendered views strip HTML comments, and templates carry much of their instruction inside comments, so a rule read from a rendered template can come out inverted.

\textbf{Identifiers.} Rule identifiers take the repository's SWE-bench instance prefix, upper-cased, and run sequentially across the whole repository. The prefix is fixed by the evaluation harness, which selects a repository's checks by it.

\subsection{Step 1: Sourcing Policy Documents}

The first stage decides which documents are in scope and produces no rules at all.

Every page is retrieved before it is assessed. Judging a page from its navigation title, or from prior familiarity with the project, is not permitted, and a page that fails to retrieve is recorded as such while the run continues. Nothing may be asserted about a file that was not read.

The scope test is applied literally to every page and section: \textit{could a rule here be violated by the diff, the commit message, or the files the agent writes?} In scope are coding style, testing, documentation, workflow, commit and pull request conventions, deprecation, and policy on automated contribution. Traversal walks the documentation navigation and then takes a single hop out of in-scope sections into the project's own pages or those of sibling repositories under the same organisation, without recursing further. Configuration files reached this way are consulted but yield no rules, source files are named but not retrieved, and an external standard such as PEP~8 yields a single rule pointing to it. Files addressed to coding agents are admitted wherever they exist, even outside the documentation navigation.

Each source is assigned one of three roles. A source carrying obligations yields rules. A source that yields no rules but makes another source decidable, such as a legend defining what each changelog fragment type means, is marked context-only and consulted. A source outside scope is excluded with a reason code recording why, covering installation and environment setup, governance, maintainer-only activity such as releasing and merging, reviewing other people's contributions, requirements on bug reports rather than contributions, and narrative material carrying no obligation.

Requirements on bug reports are kept distinct from narrative material. They are real obligations that our setting cannot reach, and folding them into narrative would assert that they were never rules.

Pages are not treated as all-or-nothing. Where a single page carries both in-scope and out-of-scope material it is marked partial and the section headings on each side are named. This is the common case rather than the exception, since repositories routinely place contributor workflow and maintainer guidance under one heading.

The reviewer confirms the manifest against Checkpoint~1 before extraction starts. Admitting maintainer pages produces a large population of rules that our setting cannot observe, which makes the evaluation setting appear more constrained than it is.

\subsection{Step 2: Extracting Atomic Policies}

Sources are worked one at a time, largest first, with the manifest row restated before each. Off-navigation sources are extracted from the raw bundle rather than from any rendered page. On a partial page, rules are drawn only from the in-scope sections, and the remainder is read for context where it governs an in-scope rule but yields no rows. Text found inside a retrieved page is treated as material to extract, never as an instruction to follow.

A rule is a repository-specific requirement, prohibition, or recommendation about creating, validating, or submitting a contribution. Description, rationale, project history, and guidance on reviewing another person's work are not rules.

Each row carries one condition and one behaviour, the sentence quoted verbatim, an atomic restatement of it, and the URL and section it came from. Bundled obligations are split, and conditions stay attached to the rule they govern rather than being separated from it. The agent also records the requirement level the modal verb alone implies, which is preserved through classification. Each rule is assigned to one of eight shared categories from a closed list, so that categories are comparable across repositories.

Two fields at this stage are load-bearing later.

\textbf{Section context} records the surrounding text that grants an exception, states a consequence, names enforcement tooling, or offers an alternative, quoted directly. The requirement level turns on it. Near-identical clauses in different repositories receive different levels depending on whether the surrounding prose sanctions deviation, and the clause alone cannot decide.

\textbf{Trigger conditions} are written into the notes in plain form for any rule conditional on the contribution touching a particular subsystem. Without them, a rule whose applicability turns on an external condition cannot be told apart from an in-scope rule that the agent's contribution simply never exercised.

Interface specifications require care, because an API inventory is not instruction prose. A page stating that a particular protocol exposes particular members is description. The rule it implies is the conformance obligation that an implementation of the protocol must expose those members, and it is extracted once, at the level of the protocol. A row that restates an individual member is cataloguing an interface rather than extracting a rule.

An in-scope section that yields no rules is recorded as having yielded none, with the reason. A source that produced nothing and a source that was overlooked are otherwise indistinguishable in the finished sheet.

When every source is done, the agent reports the final row count and identifier range, identifier contiguity, exact-duplicate quoted text across sources, rows per source including zero counts, and the distributions of modal-verb level, target, and category. The reviewer confirms the rows against Checkpoint 2, cross-referencing every extracted rule with its source sentence on the live page, with particular attention to quotations that begin mid-sentence, since a rule's real scope frequently sits in the clause immediately preceding the quoted fragment. The extraction output is then frozen as a separate file and never edited afterwards.

\subsection{Step 3: Filtering Checkable Policies}

Classification takes the frozen extraction output and applies the rubrics of Appendix~\ref{app:rubrics} along the three criteria of Section~\ref{sec:checking}: inspectable evidence, contributor actions, and mandatory requirements. For each rule the agent assigns, in a fixed order, the Evidence Type, whether the rule is in scope or the reason it is not, and for in-scope rules the requirement level (\textit{must} or \textit{should}) and the rubric route that decided it, followed by a line of reasoning. Assigning Evidence Type first is deliberate, since assigning it after scope allowed judgment rules to be admitted.

Before classification starts, the tiebreaker for conflicts within a repository is fixed centrally rather than left to the agent. Where two statements give one obligation at different levels, the stricter governs and the weaker is labelled a duplicate, with the conflict named in its notes, since an agent satisfying the stricter version satisfies both. Resolutions decided per batch come out inconsistent across the sheet.

Rules are classified in batches of 25 drawn from the whole sheet rather than page by page. Duplicate detection requires seeing rules from different sources side by side, and worked page by page, a rule stating one threshold is never recognised as duplicating a rule elsewhere that states a different one for the same obligation. After each batch the agent reports label counts and the number of rows whose rubric level differs from the modal-verb level. The two are never reconciled, since the gap between them is itself a result.

The final corpus has eleven columns. The deciding route, the modal-verb level, the section context, and the reasoning are composed into a single notes field in a fixed, delimiter-separated order, so that they remain machine-readable. Reasoning quotes the rule, states what the surrounding section granted or withheld, and gives a conclusion. Reasoning that could be copied unchanged between two rows indicates either that they are duplicates or that the agent has fallen back on boilerplate, and boilerplate makes a finished sheet impossible to audit. Where the agent could not settle a row, it still receives a label and is marked low confidence rather than left blank.

\subsection{Step 4: Compiling Policies into Checker Functions}
\label{app:checkers}
This step compiles each mandatory policy into an executable check. We describe how the checks were written, what each check ships with, and how a check decides which code it grades.

\subsubsection{Construction}

Checks were written by Claude Opus~5 at high reasoning effort in Claude Code, one
agent per project, between August and September 2026. Every agent followed the checker
authoring guide of Appendix~\ref{app:prompt-checkers}, which is released with the
benchmark. A check is written from the rule row alone: the quoted source sentence, the
atomic restatement, the evidence type, and any trigger condition recorded at extraction.
All 823 mandatory in-scope rules compiled into a checker. The 166 in-scope rules labelled
\textit{should} receive no checker by design (Section~\ref{sec:checking}).

The precondition and post-condition are described in Section~\ref{sec:checking}. Two properties of that split are worth stating explicitly here. A precondition that returns no targets yields \textit{inapplicable} rather than a pass, so a rule the agent's work never brought into scope is never counted as followed. And because the precondition is what the triggering rate of Section~\ref{sec:metrics} is computed from, an over-broad precondition inflates triggering and an over-narrow one hides violations, which is why the precondition is reviewed as a separate criterion below.

\subsubsection{Authored test cases}

Checks ship with authored cases: one that satisfies the rule, one that violates it,
and one where it does not apply (Section~\ref{sec:checking}). The cases are part of the
released artifact. They pin down how each rule is read, in a form a reader can check and
dispute.

The suite holds 2{,}603 authored cases, and 767 of the 823 checker functions (93\%) carry all three case types. The remaining 56 cannot reach all three outcomes by design. Eighteen address a human contributor in the loop, such as a certification that a person read and understood the change, so their cases cover the violating and inapplicable behaviour. Twenty-nine need evidence the bundle does not carry, such as a full suite run, so their cases show the check withholding a verdict. Nine select only the violating construct, such as a direct pytest call where SymPy provides its own wrapper, so a compliant contribution leaves them inapplicable.

\subsubsection{Exact and approximate checks}
\label{app:exact}

Each check is labelled exact or approximate. An exact check decides its policy directly, and an approximate check encodes a policy whose documentation leaves part of the reading open. Of the 823 checks, 294 are exact and 529 approximate, ranging from 24\% of SymPy's policies to 85\% of astropy's (Table~\ref{tab:corpora}). The label is assigned when the check is written, confirmed during review, and released per policy.

Restricted to the 294 exact checks, compliance rates change by at most 4.7 percentage points (Table~\ref{tab:exact}). Consolidated policies still raise compliance for every model on both scaffolds, and by more than over all checks. The model ranking is less stable: DeepSeek V4 Flash stays first in three of four columns, and the other three models reorder within a few points.

\begin{table}[H]
\centering
\caption{Compliance rate (\%) over all 823 checks and over the 294 exact checks. $\Delta$ is the gain from consolidated policies.}
\label{tab:exact}
\small
\setlength{\tabcolsep}{5pt}
\begin{tabular}{ll cc cc cc}
\toprule
 & & \multicolumn{2}{c}{Native} & \multicolumn{2}{c}{Consolidated} & \multicolumn{2}{c}{$\Delta$} \\
\cmidrule(lr){3-4}\cmidrule(lr){5-6}\cmidrule(lr){7-8}
Scaffold & Model & All & Exact & All & Exact & All & Exact \\
\midrule
\multirow{4}{*}{mini-SWE-agent}
 & GPT-5.6 Luna      & 51.8 & 52.0 & 67.6 & 69.9 & 15.8 & 17.9 \\
 & Gemini 3.7 Flash  & 53.9 & 50.7 & 65.3 & 69.0 & 11.4 & 18.3 \\
 & DeepSeek V4 Flash & 61.3 & 56.6 & 67.7 & 66.1 &  6.4 &  9.5 \\
 & Kimi K2.5         & 55.9 & 51.8 & 56.9 & 55.2 &  1.0 &  3.4 \\
\midrule
\multirow{4}{*}{OpenHands}
 & GPT-5.6 Luna      & 55.3 & 55.8 & 67.5 & 69.2 & 12.2 & 13.4 \\
 & Gemini 3.7 Flash  & 56.3 & 53.4 & 66.8 & 69.9 & 10.5 & 16.5 \\
 & DeepSeek V4 Flash & 63.8 & 59.2 & 71.2 & 70.9 &  7.4 & 11.7 \\
 & Kimi K2.5         & 57.2 & 53.4 & 62.4 & 64.6 &  5.2 & 11.2 \\
\bottomrule
\end{tabular}
\end{table}

The distinction is not a confidence score. An exact check decides its rule, as with a limit on the length of a commit summary. An approximate check stands in for a reading the documentation does not fully specify, as with a rule asking for the past tense, which the check approximates by looking for past-tense verb forms. An approximate check can be wrong in a way an exact one cannot, by disagreeing with a reader who applies the same rule, which is why the review criteria below ask separately whether the code encodes the rule and whether the label on it is right.

Eighteen checks address a human contributor in the loop, most of them from AI-assisted contribution policies, such as a requirement that a person certifies they read and understood the change. We score them like every other check, since the projects publish these policies for any contributor, autonomous agents included. Each carries a flag in the release, so rates can also be computed without them.

\subsubsection{How an approximate check decides}
\label{app:approx}

A check calls no model, so a policy without an exact test must still reduce to plain code. An approximate check tests a visible trace of the property the policy asks for, such as a word, a code construct or a file location. The trace can stand in for the situation the policy covers, in the precondition, or for the required behaviour, in the post-condition. The docstring of every approximate check names the trace and the direction in which it can err. We walk through one check in full, then list other techniques in Table~\ref{tab:approx-techniques}.

\paragraph{The policy.} Django asks contributors to \textit{phrase commit subject lines in past tense and end them with a period} (DJANGO-C046). One of our agents wrote the compliant subject \textit{Fixed \#29528 -{}- Made URLValidator reject invalid characters in the username and password.} Most projects ask for the opposite, an imperative subject with no period. A check written from habit would grade Django backwards, so the check follows the corpus sentence.

\paragraph{The check.} The precondition selects every commit the agent made. The post-condition grades two clauses separately, so the reason for a violation names the clause that failed (Figure~\ref{fig:c046-code}). The period clause is exact. The tense clause is approximate, since no short rule decides English tense. The check reads the leading verb of the subject, which is the first word, or the first word after Django's ticket prefix \textit{Fixed \#N -{}-}. A verb counts as past tense if it ends in \textit{-ed} or appears in a list of 118 irregular past forms, such as \textit{made}, \textit{kept} and \textit{wrote}. Either leading verb in the past tense satisfies the clause.

\begin{figure}[h]
\begin{tcolorbox}[colback=gray!4, colframe=gray!45, boxrule=0.4pt, left=4pt, right=4pt]
\begin{Verbatim}[fontsize=\scriptsize, breaklines=true]
_IRREGULAR_PAST = frozenset("made took began broke ... wrote withdrew".split())

def _looks_past_tense(word):
    w = word.lower()
    return w.endswith("ed") or w in _IRREGULAR_PAST

@rule(id="DJANGO-C046", category="Git and commit conventions",
      ownership="created", heuristic=True, reads=("commits",))
class PastTenseSubjectWithPeriod:
    def precondition(self, bundle):                  # every commit the agent made
        return [commit_target(c) for c in bundle.commits]

    def post_condition(self, target):
        summary = target.payload.summary.strip()
        problems = []
        if not summary.endswith("."):                                    # exact
            problems.append("subject does not end with a period")
        if not any(_looks_past_tense(w) for w in _leading_verbs(summary)):  # approximate
            problems.append("subject is not phrased in the past tense")
        return Violated("; ".join(problems)) if problems else Satisfied(summary)
\end{Verbatim}
\end{tcolorbox}
\caption{The check for DJANGO-C046, trimmed. \texttt{\_leading\_verbs} returns the first word of the subject and the first word after the ticket prefix. The full list holds 118 irregular forms.}
\label{fig:c046-code}
\end{figure}

\paragraph{Authored cases.} Five cases pin this reading (Table~\ref{tab:c046-cases}). The two violating cases each break one clause, so each clause is tested on its own.

\begin{table}[h]
\centering
\small
\caption{Authored cases for DJANGO-C046.}
\label{tab:c046-cases}
\begin{tabular}{@{}lll@{}}
\toprule
Commit subject & Outcome & Reason \\
\midrule
\texttt{Fixed \#123 -{}- Added truncation to Truncator.} & satisfied & \\
\texttt{Made Truncator keep HTML entities.} & satisfied & irregular verb \\
\texttt{Add truncation to Truncator.} & violated & not in the past tense \\
\texttt{Added truncation to Truncator} & violated & no final period \\
(no commit) & inapplicable & \\
\bottomrule
\end{tabular}
\end{table}

\paragraph{Agreement with a reader.} We ran the check on all 3,685 commits in our 3,696 Django runs and read the leading verb of each subject. Setting aside 24 commits whose verb (\textit{set}, \textit{reset}, \textit{unset}) is spelled the same in both tenses, the check agrees with the reader on 3,647 of 3,661 (99.6\%). Thirteen of the 14 disagreements accept the verb in the \textit{Fixed \#N} prefix while the description after it is imperative. The other is \textit{Doc'd} (short for \textit{documented}), a past form missing from the list.

\paragraph{Other techniques.} The past-tense check uses a word list. Table~\ref{tab:approx-techniques} gives four other techniques, with one policy each. Every technique has the same limit: the check errs when the trace and the property come apart, and the last column names that case. The human audit of Appendix~\ref{app:checker-review} covers 97 approximate checks.

\begin{table}[h]
\centering
\small
\caption{Techniques used by approximate checks, with one policy each.}
\label{tab:approx-techniques}
\begin{tabularx}{\linewidth}{@{}>{\raggedright\arraybackslash}p{1.9cm}>{\raggedright\arraybackslash}p{3.2cm}>{\raggedright\arraybackslash}X>{\raggedright\arraybackslash}X@{}}
\toprule
Technique & Policy & What the check tests & Case it gets wrong \\
\midrule
Word list & DJANGO-C046: commit subjects in the past tense & the leading verb ends in \textit{-ed} or is a listed irregular form & misses past forms outside the list \\
Text pattern & ASTROPY-C153: \texttt{astropy} for the package, Astropy for the Project & each mention matches one of the two sanctioned spellings & cannot tell which meaning a sentence intends \\
Code structure & SCIKIT-LEARN-C158: estimators inherit from \texttt{BaseEstimator} & a new class that defines \texttt{fit} is an estimator & selects other classes with a \texttt{fit} method \\
File location & MATPLOTLIB-C262: imported code carries a compatible licence & a new file under \texttt{extern/} is imported code & misses imported code placed elsewhere \\
Co-change & SCIKIT-LEARN-C118: deprecations are listed in the API reference & a \texttt{@deprecated} name comes with an edit to the reference list & does not check that the right name was added \\
\bottomrule
\end{tabularx}
\end{table}

\subsubsection{Ownership and determinism}
\label{app:ownership}

A check grades only code the agent is responsible for, never pre-existing repository content. Each check declares one of three ownership modes. \emph{Created} applies to existence rules, which ask whether the agent produced an artifact at all. \emph{Touched} applies to formatting rules, which are graded only on the lines the diff changed. \emph{Enclosing} applies to consistency rules: editing a function makes the agent responsible for its docstring and doctest, but not for prose it did not write. A wider reading would grade the target repository's own authors, who are not under test (Appendix~\ref{app:reachability}).

Checks are ordinary Python code, written once and run unchanged on every stored run. Scoring a run writes no code and calls no model, so the same run always receives the same verdicts, and a test enforces this. We considered grading each rule with a model as judge~\citep{zheng2023judging} and rejected it for two reasons. The same run could score differently when graded again, and one model's compliance would be graded by another model's judgment. Where a rule names a static tool that runs on the changed files alone, such as flake8 or black, the check runs that tool on the base and the head and scores the difference.

\subsection{Step 5: Validating Checker Functions}
\label{app:validation}

\subsubsection{Review procedure}
\label{app:checker-review}

The checks are generated code, so they are reviewed before release rather than trusted. Review has three parts, an automated pass over the whole population, execution against the authored cases, a blind human audit of a stratified sample. The instrument the human raters worked from is reproduced in full in Appendix~\ref{app:review-guidelines}, since the guidelines are what the agreement figure is agreement about.

\textbf{Automated pass.} Every check is reviewed by Claude Opus 5 in Claude Code against the same criteria the human raters use, yielding one verdict and one line of reasoning per check. This pass locates likely faults and defines the strata the human sample is drawn from. It is not reported as human validation, and no released check rests on it alone.

\textbf{Execution.} Reading code detects a check that encodes the wrong rule but not one that encodes the right rule incorrectly, so every check is executed against its authored cases and the verdicts are recorded with the item. A check whose cases come out wrong is corrected before the sample is drawn. No check in the sample failed a case it carries. Where a sampled check has no case of a given type, the rater tests that behaviour on a patch of their own construction.

\textbf{Sample.} Two raters independently audit a stratified sample of 150 checks, or 18\% of the 823, drawn by a released script under a recorded seed. Allocation is not proportional, because corpus size ranges from 2 to 167 checks and proportional allocation would leave the smallest projects unrepresented. Projects with ten or fewer checks are taken in full, every remaining project receives a floor of five, and the remaining budget is distributed in proportion to corpus size, with at least one differential and one trajectory check drawn from every project that has any. Table~\ref{tab:audit-sample} gives the per-project counts. The sample carries 53 exact checks and 97 approximate ones, against 294 and 529 in the population, and 119 output, 14 differential and 17 trajectory checks. Because small projects are deliberately overrepresented, any population-level rate computed from the sample is weighted by each project's share of the 823.

For each check, and without sight of the automated verdict or of the other rater's labels, a rater answers whether the code encodes the rule as extracted, neither more strictly nor more loosely than the quoted source sentence; whether the precondition selects the targets the rule applies to, and in particular whether a recorded trigger condition is what it tests; whether the post-condition fires on the violating case and stays silent on the satisfying one; and whether the check is exact or approximate in the sense of Appendix~\ref{app:exact}. The rater then records one verdict, \textit{accept} or \textit{reject}, naming the failing criterion and the fault in the case of a rejection. The verdict is binary by design. Whether a faulty check can be corrected with a small edit or has to be removed is a decision about the remedy rather than about the code, and folding it into the label would have the two raters disagreeing over repairability rather than over correctness. A rater who cannot settle an item records a rejection, since \textit{accept} asserts that the check is correct.

\begin{table}[h]
\centering
\caption{Composition of the checker audit sample. Projects with ten or fewer checks are censused rather than sampled. \textit{Diff.} and \textit{traj.} give the differential and trajectory checks drawn, the two evidence types that chance alone would leave out.}
\label{tab:audit-sample}
\small
\begin{tabular}{lrrrr}
\toprule
project & checks & drawn & diff. & traj. \\
\midrule
matplotlib   & 167 & 24 & 2 & 1 \\
scikit-learn & 146 & 21 & 1 & 1 \\
SymPy        & 142 & 21 & 1 & 3 \\
astropy      & 111 & 18 & 1 & 1 \\
Django       &  78 & 14 & 2 & 1 \\
pylint       &  49 & 11 & 1 & 1 \\
xarray       &  35 &  9 & 1 & 1 \\
Flask        &  34 &  9 & 2 & 1 \\
Sphinx       &  24 &  8 & 1 & 2 \\
pytest       &  22 &  7 & 1 & 2 \\
Requests     &  13 &  6 & 1 & 2 \\
seaborn      &   2 &  2 & 0 & 1 \\
\midrule
total        & 823 & 150 & 14 & 17 \\
\bottomrule
\end{tabular}
\end{table}

The audit was carried out by two annotators with software engineering experience.
Neither annotator wrote or reviewed any checker before the audit. Each worked
independently, without sight of the automated verdicts or of the other annotator's
labels.

\subsubsection{Agreement}

Table~\ref{tab:agreement} reports agreement between the two raters on the 150 audited checks, as raw percentage agreement and Cohen's $\kappa$ \citep{cohen1960coefficient}. The raters agreed on 141 of the 150 verdicts (94.0\%).

\begin{table}[h]
\centering
\caption{Agreement between the two raters on the 150 audited checks, on the verdict and on each of the four questions.}
\label{tab:agreement}
\small
\begin{tabular}{lcc}
\toprule
 & raw agr. (\%) & Cohen's $\kappa$ \\
\midrule
verdict (accept / reject)  & 94.0 & 0.72 \\
C1 encodes the rule        & 99.3 & 0.89 \\
C2 precondition            & 96.0 & 0.73 \\
C3 post-condition          & 98.0 & 0.72 \\
C4 exact or approximate    & 96.7 & 0.93 \\
\bottomrule
\end{tabular}
\end{table}

Raw agreement is at or above 94.0\% on the verdict and on all four questions, and $\kappa$ is substantial throughout (0.72--0.93) against skewed marginals that make $\kappa$ the conservative statistic here. The two raters therefore apply the same standard, which is the condition under which the sample estimates the population.

The raters differ on 10 items across the verdict and the three pass criteria. Six are C2, and all six concern a precondition that selects more than its rule covers. C1 and C3 are near-unanimous, since each is judged against fixed material: the quoted source sentence for C1, the authored cases for C3.

\textbf{Outcome.} The raters accepted 135 and 128 of the 150 checks, and 127 were accepted by both. Weighted by each project's share of the 823, the mean rater accept rate is 87.2\% (95\% CI [82.1, 92.3], stratified by project with a finite population correction); requiring both raters to accept gives 83.7\%. Weighting matters because the sample overrepresents small projects by design. No check was corrected or removed after the audit, so the released corpus is the audited corpus.

\textbf{What the audit supports.} Two raters who agree at 94.0\% accept 87.2\% of a seeded stratified sample, so an estimated 87\% of the 823 checks encode their rule, with the interval above. The audit bounds the corpus and makes no claim about an unsampled check. The residual is concentrated in over-selecting preconditions. The raters do not maintain these projects, so the audit is not maintainer judgment.

\subsection{Acceptance Checks and Failure Modes}
\label{app:acceptance}

A repository is accepted only after passing the three checkpoints in Table~\ref{tab:checkpoints}. An author performs the first two, and an acceptance script performs the third.

\begin{table}[h]
\caption{Checkpoints in the extraction workflow.}
\label{tab:checkpoints}
\begin{center}
\small
\begin{tabularx}{\linewidth}{lX}
\multicolumn{1}{c}{\bf Checkpoint} & \multicolumn{1}{c}{\bf Checks} \\
\hline \\
1. Manifest (author) & every URL is at the pinned version; no maintainer or triage page admitted whole; partial pages name their in-scope sections; configuration files are context-only; nothing is claimed about a file that was not retrieved \\
2. Extraction (author) & row count plausible against the other repositories; section context and notes populated on every row; every rule records its source sentence and URL; identifiers contiguous; every category from the closed list; every rule cross-referenced with its source sentence on the live page; mid-sentence quotations checked against the preceding clause \\
3. Acceptance (automated) & rules-to-source-length ratio within band; judgment and in scope never co-occur; modal-verb and rubric levels differ on some rules, with direction recorded; reasoning distinct across rows; not-observable counted separately from other out-of-scope reasons; schema conformance \\
\end{tabularx}
\end{center}
\end{table}

\paragraph{Failure modes.}

The following were observed during construction. Most are specific to model-driven extraction and would not arise if the documentation were read by hand. Each is addressed by a step above.

\begin{table}[h]
\caption{Failure modes observed during rule extraction.}
\label{tab:failure-modes}
\begin{center}
\small
\begin{tabularx}{\linewidth}{XXX}
\multicolumn{1}{c}{\bf Symptom} & \multicolumn{1}{c}{\bf Cause} & \multicolumn{1}{c}{\bf Mitigation} \\
\hline \\
Version drift mid-run & stable documentation retrieved & pin the developer documentation \\
False version mismatch & pinned at build rather than release & halt only on a different release \\
Template rules missing or inverted & rendered view strips comments & retrieve raw text in the pre-pass \\
Configuration files never retrieved & not reachable from navigation & resolve by path in the pre-pass \\
Agent-directed policy missing & agent files outside navigation & admit them in traversal and the pre-pass \\
Rules asserted unread & model answers from prior familiarity & require retrieval or an explicit non-retrieval \\
Instructions lost mid-run & context fills over a long run & restate the manifest row per source \\
Quotation starts mid-sentence & clause boundary misjudged & check the preceding clause at review \\
Interface pages inflate counts & cataloguing an API & extract the conformance obligation, not members \\
Model follows instructions in a page & injected text in documentation & extract from it, never obey it \\
Modal-verb level lost & only the final level written & modal-verb level required on every row \\
Out-of-scope reasons pooled & reason written without its code & reason field must carry its code \\
Invented categories & category list treated as open & closed list of eight \\
Judgment rules admitted & Evidence Type assigned after scope & Evidence Type assigned first \\
Repository scores 0 of 0 & shortened identifier prefix & prefix fixed by the harness \\
\end{tabularx}
\end{center}
\end{table}

\section{Experiment Details and Additional Results}
\label{app:results}

This appendix supports Section~\ref{sec:setup} and the results that follow it. The prompts given to the agents are in Appendix~\ref{app:agent-prompts}.

\subsection{Run Coverage}
\label{app:coverage}

Compliance grading reads the stored runs offline, so it covers all 8{,}000 runs (500 instances, four models, two settings, two scaffolds). Functional grading runs the SWE-bench harness and returns no verdict on 61 runs (0.8\%). None of these failures comes from the compliance checkers. Most are agent outcomes: the agent submitted an empty patch, reached the 3{,}600\,s run cap, was stopped by the OpenHands loop detector, or left build output in its patch, which then clashes with the SWE-bench test patch. The rest are model provider errors and environment setup failures (image pull, image build, container start) that persisted after retries. We do not edit a stored patch to make it gradable. As in SWE-bench, a run without a verdict counts as unresolved.

\begin{figure}[h]
\centering
\includegraphics[width=0.85\linewidth]{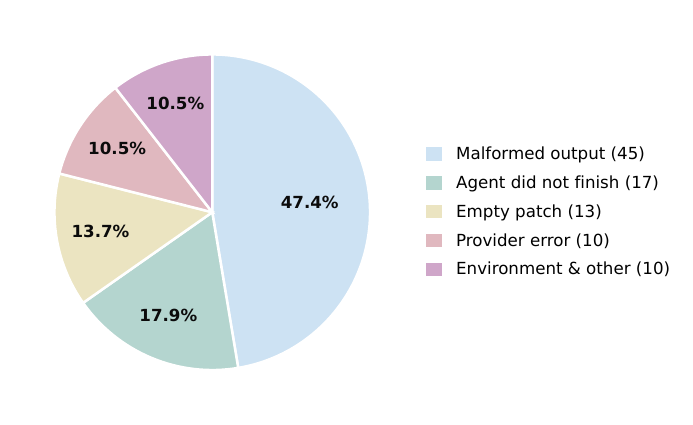}
\caption{Runs with an evaluation gap, by cause, over all 8{,}000 runs. \textit{No functional verdict}: the SWE-bench harness returns no result. \textit{Compliance partly withheld}: the run is graded, but checkers that need a parsed file return no verdict.}
\label{fig:run-coverage}
\end{figure}

Separately, in 36 runs a file the agent changed does not parse as Python. A checker that needs the file's syntax tree then returns no verdict, and we do not repair the file, since that would add work the agent did not do. The affected policies count as withheld (Appendix~\ref{app:hazards}). Together the two groups cover 95 distinct runs, since 2 runs fall in both. Figure~\ref{fig:run-coverage} gives their causes.

\subsection{Comparison with Public Resolve Rates}
\label{app:vals}

Table~\ref{tab:vals} compares our Native resolve rate under mini-SWE-agent with the public SWE-bench Verified results of Vals AI, which use the same scaffold~\citep{vals2026swebench}. Three of the four models fall within 5 percentage points of the public result. GPT-5.6 Luna resolves 14.4 points fewer instances. Vals AI runs OpenAI models through the scaffold's Responses API model class, while we keep the default model class for all four models, so that every model runs under the same unmodified mini-SWE-agent configuration.

\begin{table}[h]
\centering
\caption{Resolve rate (\%) on SWE-bench Verified under mini-SWE-agent.}
\label{tab:vals}
\begin{tabular}{lrrr}
\toprule
Model & Ours (Native) & Vals AI & $\Delta$ \\
\midrule
GPT-5.6 Luna           & 78.6 & 93.0 & $-14.4$ \\
Gemini 3.7 Flash       & 80.2 & 80.8 & $-0.6$ \\
DeepSeek V4 Flash 0731 & 93.4 & 88.8 & $+4.6$ \\
Kimi K2.5              & 74.6 & 70.0 & $+4.6$ \\
\bottomrule
\end{tabular}
\end{table}

\subsection{Policy Coverage}
\label{app:triggering}
\label{app:reachability}

The rule corpus states everything a project asks of a contributor, while SWE-bench Verified exercises one narrow kind of contribution, a bug fix. The two overlap only in part. Of the 823 policies, 304 never trigger on any of the 8{,}000 runs (Table~\ref{tab:reachability}). Their silence reflects the pairing of corpus and task set, and says nothing about whether agents comply.

A policy that triggers is graded only when its checker finds the evidence it needs in the run. Table~\ref{tab:bycat-triggering} gives both counts per category. The gap is small in most categories and large in two. In \textit{Code and quality}, 27\% of triggered policies are graded, and in \textit{Tests and test style}, 64\%. Most of these policies need a type checker or a test run in the full project environment to decide, such as ``a new test fails on the code before the fix'', and a stored run holds no such result. The checker then withholds a verdict (Appendix~\ref{app:hazards}). The compliance rates for these two categories in Table~\ref{tab:bycat} therefore rest on the graded subset.

\begin{table}[h]
\centering
\small
\caption{Policies that never trigger on any of the 8{,}000 runs, per category.}
\label{tab:reachability}
\begin{tabular}{lrrr}
\toprule
Category & Policies & Never triggered & Share \\
\midrule
Specialized changes             & 107 &  81 & 76\% \\
Documentation and docstrings    & 224 &  93 & 42\% \\
Language and framework style    & 163 &  59 & 36\% \\
Tests and test style            & 160 &  50 & 31\% \\
Git and commit conventions      &  40 &  10 & 25\% \\
Code and quality                &  35 &   5 & 14\% \\
PR and release metadata         &  57 &   6 & 11\% \\
AI-assisted contribution policy &  37 &   0 &  0\% \\
\midrule
Total                           & 823 & 304 & 37\% \\
\bottomrule
\end{tabular}
\end{table}

% GENERATED by `python tools/table2.py` -> output/table2.tex (second table); rate columns dropped.
\begin{table}[t]
\centering
\caption{Policies triggered and graded per category, pooled over twelve projects and four models within each scaffold. The gap is the policies whose checker finds too little evidence in the run to return a verdict.}
\label{tab:bycat-triggering}
\small
\setlength{\tabcolsep}{4pt}
\begin{tabular}{llrrrr}
\toprule
 & & \multicolumn{2}{c}{\textbf{Native}} & \multicolumn{2}{c}{\textbf{Consolidated}} \\
\cmidrule(lr){3-4}\cmidrule(lr){5-6}
Category & Scaffold & Triggered & Graded & Triggered & Graded \\
\midrule
Git and commit conventions & mini-SWE-agent & 4{,}100 & 4{,}092 & 4{,}676 & 4{,}671 \\
 & OpenHands & 4{,}361 & 4{,}353 & 4{,}730 & 4{,}722 \\
\addlinespace
PR and release metadata & mini-SWE-agent & 4{,}619 & 4{,}619 & 5{,}793 & 5{,}793 \\
 & OpenHands & 4{,}764 & 4{,}764 & 6{,}586 & 6{,}586 \\
\addlinespace
Code and quality & mini-SWE-agent & 5{,}385 & 1{,}453 & 5{,}388 & 1{,}484 \\
 & OpenHands & 5{,}304 & 1{,}462 & 5{,}327 & 1{,}484 \\
\addlinespace
AI-assisted contribution policy & mini-SWE-agent & 7{,}381 & 7{,}380 & 7{,}477 & 7{,}475 \\
 & OpenHands & 7{,}508 & 7{,}507 & 7{,}733 & 7{,}732 \\
\addlinespace
Tests and test style & mini-SWE-agent & 6{,}054 & 3{,}872 & 6{,}110 & 3{,}956 \\
 & OpenHands & 6{,}031 & 3{,}886 & 6{,}151 & 4{,}002 \\
\addlinespace
Language and framework style & mini-SWE-agent & 12{,}897 & 11{,}732 & 12{,}941 & 11{,}776 \\
 & OpenHands & 13{,}114 & 11{,}956 & 13{,}188 & 12{,}026 \\
\addlinespace
Specialized changes & mini-SWE-agent & 278 & 268 & 265 & 256 \\
 & OpenHands & 259 & 250 & 274 & 264 \\
\addlinespace
Documentation and docstrings & mini-SWE-agent & 2{,}466 & 2{,}287 & 2{,}835 & 2{,}630 \\
 & OpenHands & 2{,}727 & 2{,}505 & 3{,}098 & 2{,}868 \\
\bottomrule
\end{tabular}
\end{table}

\paragraph{Specialized changes.} Specialized changes is the least reachable category: 81 of its 107 policies
never trigger. Most of them encode procedures a bug fix does not involve.
SymPy's 27 policies in the category show the pattern. Most describe one
procedure end to end: introducing a deprecation requires setting
\texttt{deprecated\_since\_version} and \texttt{active\_deprecations\_target},
passing \texttt{stacklevel}, writing the deprecation message as one plain-text
migration paragraph, annotating the docstring, adding a section to
\texttt{active-deprecations.md}, adding a test that both triggers the warning
and checks the deprecated behavior still holds, and flagging the change
\texttt{BREAKING CHANGE} in the release notes. SWE-bench Verified instances are
selected for a clear, test-verifiable bug fix. Deprecating an API is different
work, and usually its own pull request. Only 2 of SymPy's 27 policies trigger
in the full sweep, both on changes that break backwards compatibility
(SYMPY-C239 and SYMPY-C268).

\paragraph{Documentation and docstrings.} Documentation and docstrings holds the most policies, 224, yet triggers in
only 5 to 7\% of cells (Table~\ref{tab:bycat}). The cause is ownership. A
documentation policy is graded only on documentation the agent writes or
edits (Appendix~\ref{app:ownership}). An agent that fixes a function without
touching its docstring brings no documentation policy into scope, and grading
it on the form of a docstring it did not edit would score the project's own
authors. The category is reachable: 31 of SymPy's 40 documentation policies
trigger at least once in the full sweep. Agents rarely do the work that
reaches it.

\subsection{Triggered and Withheld Policies}
\label{app:hazards}

A compliance rate alone can reward an agent for doing less work, since fewer actions trigger fewer policies. A one-line commit with no pull request text triggers 6 policies and passes all 6 (100\%), while a submission with a commit body, a pull request description and a release-notes entry triggers 24 and passes 23 (96\%). The same effect appears in the full sweep: a Consolidated run triggers 22.7 policies on average under mini-SWE-agent and 23.5 under OpenHands, against 21.6 and 22.0 for a Native run. We therefore report the triggering rate alongside the compliance rate (Section~3.5):
\[
\text{triggering rate} = \frac{\text{policies triggered}}{\text{policies in the repository}}, \qquad
\text{compliance rate} = \frac{\text{policies passed}}{\text{policies triggered}}.
\]
A triggered policy is withheld when the evidence it needs is missing from the run, such as a file that does not parse or a property that requires a test-suite run. Withheld policies count as triggered and are excluded from the compliance rate, so a limit of the checker is reported neither as less work by the agent nor as a verdict on it.

\section{Rubrics and Prompts}
\label{app:rubrics-prompts}

This appendix reproduces the instruments behind Appendix~\ref{app:pipeline}: the two labelling rubrics used in Step~3, the guidelines the human reviewers followed in Step~5, the prompts for benchmark construction, and the prompts given to the evaluated agents.

\subsection{Labelling Rubrics}
\label{app:rubrics}

Each rule is a single atomic obligation extracted verbatim from a repository's contribution documentation and restated as one imperative sentence; the extraction procedure is described in Appendix~\ref{app:pipeline}.

A rule enters the benchmark only if it meets three criteria, and each criterion fills one field of the released dataset. \textit{Inspectable evidence}: the rule names an artifact a checker can open, recorded in the Evidence Type field. \textit{Contributor actions}: the run produces that artifact and the rule governs what the agent itself does, recorded in the In scope field by Rubric A. \textit{Mandatory requirements}: the rule states a requirement, recorded as \textit{must} in the Requirement level field by Rubric B. This subsection defines Evidence Type and gives the two rubrics in full, as applied. Each rubric is a decision procedure: the routes are evaluated in a fixed order, the first one that fires decides the label, and the route that fired is recorded on the row. We state them in the form the annotators worked from, so that the labels in our released dataset can be reproduced or contested.

The criteria are applied in a fixed order. Evidence Type is assigned first, then scope, then requirement level on in-scope rules only. The order matters: the evidence a rule needs constrains whether it can be scored at all, and an unscoreable rule is never graded for requirement level.

\subsubsection{Inspectable evidence (field: Evidence Type)}

Every rule that we keep has to be checked against something the run produced. Evidence Type records what that something is. The four values answer one question: what does the checker have to open in order to decide whether the rule was followed?

\begin{table}[h]
\caption{Evidence Type.}
\label{tab:evidence-type-def}
\begin{center}
\begin{tabular}{ll}
\multicolumn{1}{c}{\bf Value} & \multicolumn{1}{c}{\bf What the check opens} \\
\hline \\
output       & the final files or the commit message, read once, with no comparison \\
differential & two states, compared: pre-patch against post-patch \\
trajectory   & a record of what the agent did, not only what it produced \\
judgment     & nothing; the standard is not written down and a human decides \\
\end{tabular}
\end{center}
\end{table}

\textbf{output.} The check reads the end state and nothing else. One pass over the files the agent wrote, or over the commit message it produced. No comparison to a previous state, no history, no record of how the work was done. Most documented contribution rules are of this kind, because most of them describe what the code should look like when the work is finished.

\textbf{differential.} The check needs two states and a comparison between them. Reading either state alone gives the wrong answer. Regression-test requirements are the clearest case: a test that passes after a fix satisfies nothing on its own, because a test that also passed before the fix is not a regression test. Rules scoped to newly added code are the same shape, since they require knowing which code is new.

\textbf{trajectory.} The check needs a record of what the agent did. Two patches can be byte-identical while one satisfies the rule and the other does not, because the rule constrains the process rather than the product. Rules conditioned on the kind of change made, and rules about disclosing how the work was produced, both fall here.

\textbf{judgment.} There is no artifact to open, because the rule does not say what passing looks like. These are real editorial standards that a human reviewer applies without difficulty and that no checker can apply at all, because the standard lives in the reviewer rather than in the documentation.

Judgment determines scope by itself: if the evidence is a human opinion, the rule is out of scope by route N2. A row marked judgment and In Scope means the Evidence Type was assigned wrongly, not that an exception was found, and we re-decide the tier when the combination appears.

\subsubsection{Contributor actions (Rubric A; field: In scope)}

Two labels, \textit{In Scope} and \textit{Out of scope}. A rule is in scope when the run produces evidence bearing on it, and we can say what that evidence would have to show. Strength is graded on in-scope rules only.

\paragraph{The test.} A rule is in scope when both of the following hold.

\begin{table}[h]
\caption{Conditions for a rule to be in scope.}
\label{tab:confirm}
\begin{center}
\begin{tabular}{lll}
\multicolumn{1}{c}{} & \multicolumn{1}{c}{\bf Condition} & \multicolumn{1}{c}{\bf What the annotator does} \\
\hline \\
C1 & observability & names the file, diff, or output that the check opens \\
C2 & decidability  & writes down the failure condition \\
\end{tabular}
\end{center}
\end{table}

C2 is the stricter of the two. If stating the failure condition requires a word such as \textit{reasonably}, \textit{usually}, or \textit{appropriate}, then the condition has not been written down.

A rule can fail this test in four ways, and they are not the same kind of failure. In one case the evidence does not exist and no amount of effort on our part would produce it. In the other three the evidence exists and we chose not to score the rule. We keep the two apart everywhere, in the labels and in every count we report, because pooling them would make a scope we narrowed by choice indistinguishable from one the setting forced on us.

\paragraph{The harness produced no evidence.}

\textit{Not observable} (N1). The rule governs an artifact that a patch-only setting never creates. There is nothing to open, so C1 fails and the question of what passing would mean does not arise.

Which rules fall here is fixed by the setting rather than by the rule, so we state the setting directly:

\begin{table}[h]
\caption{What a patch-only harness does and does not produce.}
\label{tab:harness}
\begin{center}
\begin{tabular}{ll}
\multicolumn{1}{c}{\bf Exists} & \multicolumn{1}{c}{\bf Does not exist} \\
\hline \\
repository working tree, the agent's diff & pull request object, PR template, review thread \\
one commit and its message                & multiple commits, branch history, rebase or squash \\
a local test suite run                    & CI service, coverage bot, issue tracker \\
the files the agent writes                & built or rendered documentation, browser, screenshots \\
                                          & a second human, contributor identity \\
\end{tabular}
\end{center}
\end{table}

A rule whose subject appears in the right-hand column is not observable, and this is the one route whose population would change if the harness changed.

\paragraph{The evidence exists and we did not score it.}

\textit{Subjective} (N2). The artifact is there, but the documentation does not state what counts as passing. C1 holds and C2 fails. Every rule typed as \textit{judgment} arrives here.

\textit{Duplicate} (N3). The rule is checkable, but the check would open the same artifact another rule already opens and would reach the same verdict. Scoring it twice would double-count a single observation.

\textit{Trigger} (N4). The rule is checkable and decidable, but whether it applies at all depends on a fact the run does not contain, so we cannot tell whether to score it on a given task.

\paragraph{Order of evaluation.} The four are evaluated in the order given, and the first that fires decides the label. The ordering is literal: not observable fires before subjective even where a rule is also a quality call, because if the artifact does not exist there is nothing to be subjective about. Where more than one route applies we record the first and name the others on the row.

\paragraph{Not observable against trigger.} These two are the pair most often confused, since both amount to ``we cannot score this on a given run''. They are distinguished by a single test:

\begin{quote}
Does the artifact the rule governs exist in the run?
\begin{itemize}
\item No: \textbf{not observable}.
\item Yes, but whether the rule applies turns on a fact outside the agent's own behaviour: \textbf{trigger}.
\item Yes, and whether it applies is determined by what the agent itself did: \textbf{in scope}.
\end{itemize}
\end{quote}

The load-bearing clause is \textit{outside the agent's own behaviour}. Contributor identity and first-time-contributor status sit outside it, so rules conditioned on them are triggers. Whether the agent used a particular tool sits inside it, because the run records what the agent did, so such rules stay in scope even where we can predict in advance that they will fail.

\subsubsection{Mandatory requirements (Rubric B; field: Requirement level)}

Two labels, \textit{must} and \textit{should}, graded on in-scope rules only. Prohibitions are \textit{must}. Any one route is sufficient; we record which fired.

\begin{table}[h]
\caption{Routes to \textit{must}.}
\label{tab:must}
\begin{center}
\begin{tabular}{ll}
\multicolumn{1}{c}{\bf Route} & \multicolumn{1}{c}{\bf The rule\ldots} \\
\hline \\
M1 & says so outright (\textit{must}, \textit{required}) \\
M2 & forbids something (\textit{never}, \textit{do not}, \textit{must not}) \\
M3 & is a condition of acceptance: a pre-merge or review checklist, or a named CI job \\
M4 & names an exact thing, limit, form, or ordering \\
M5 & is stated non-mandatorily, but admits only one satisfying state \\
M6 & states a consequence in the source \\
\end{tabular}
\end{center}
\end{table}

\begin{table}[h]
\caption{Routes to \textit{should}.}
\label{tab:should}
\begin{center}
\begin{tabular}{ll}
\multicolumn{1}{c}{\bf Route} & \multicolumn{1}{c}{\bf The rule\ldots} \\
\hline \\
D1 & uses preference wording \\
D2 & permits deviation in its documentation \\
D3 & has a fixed form, but whether it applies is a judgment call \\
D4 & instructs on \textit{how}, not \textit{what} \\
\end{tabular}
\end{center}
\end{table}

\paragraph{Precedence.} Where more than one route fires:

\begin{itemize}
\item \textbf{M3 beats everything.} A condition of acceptance is a gate regardless of tone.
\item \textbf{D2 beats M4 and M5.} An exception the documentation leaves open destroys the single right answer. An exception written into the rule itself does not.
\item \textbf{D3 beats M5.} The split is applicability, not wording.
\item \textbf{M5 beats D1.} Non-mandatory phrasing is not on its own a reason to demote. One named command, decorator, value from a published list, or exact format is \textit{must}.
\end{itemize}

\paragraph{Reading the section, not the clause.} Strength turns on the text surrounding a rule as much as on the rule itself. Two repositories in our set state near-identical commit-message length limits and receive different labels, because one set of documents sanctions deviation in the surrounding prose and the other does not. Annotators read the full section before assigning strength, and the sanctioning text is recorded on the row.

\paragraph{Object against manner.} A rule naming a specific token to use or avoid is \textit{must} by M4, even when phrased with a soft verb such as \textit{avoid}. A rule describing a property the result should have is \textit{should} by D4. The distinction is what the rule points at, not how firmly it is worded.

\paragraph{Conflicts within a repository.} Two rules may state one obligation at different strengths, on the same page or across pages; the precedence table settles single rules, not pairs. The stricter statement governs and the weaker is recorded as N3 with the conflict named, because an agent satisfying the strict version satisfies both. We fix this tiebreaker before classification begins rather than resolving conflicts case by case.

\paragraph{Logging.} Every in-scope row records the route that fired, M1 to M6 or D1 to D4. Free-text reasoning supplements this and does not replace it. There is no \textit{prohibited} label in the released data: prohibitions are folded into \textit{must} by M2, and the fold is recorded. Rows the annotator could not settle still receive a label, marked low confidence.
\subsection{Checker Review Guidelines}
\label{app:review-guidelines}

Both raters worked from the following document unchanged. It is reproduced verbatim, including the sheet mechanics, because the wording of the criteria is what the raters were actually asked and a reader assessing the agreement figure needs to see it rather than a summary of it.

\begin{tcolorbox}[breakable, colback=gray!4, colframe=gray!45, boxrule=0.4pt, left=4pt, right=4pt]
\begin{Verbatim}[fontsize=\scriptsize, breaklines=true, breakanywhere=true]
# Checker review guidelines

Both raters use this unchanged. One item is one compliance check.

## What you are judging

Each row gives you a rule and the code that grades it. The column names follow the
paper's terminology, so evidence type takes the values output, differential and
trajectory, and a check is labelled exact or approximate. The code has two parts, a
precondition that picks out what the rule applies to, and a post-condition applied to
each of those in turn.

You are judging whether the code is a correct encoding of that rule. You are not
judging whether the rule was worth extracting, whether it was classified correctly, or
whether the code is tidy. If you notice a problem with the rule itself, put it in the
notes and label the code as you find it.

## Before you start

Work alone. Do not look at the other rater's sheet or discuss items until both sheets
are in.

Budget about five minutes an item. If one takes more than ten, reject it, say in the
notes what you would need in order to settle it, and move on. Stop after two hours in
a sitting, since tiredness shows up as drift toward accept.

## The four questions

Answer C1 to C3 `yes` or `no`, and C4 `exact` or `approximate`. Answer all four even when
one of them fails, since which question failed is what tells us where the two of you
diverged.

**C1. Does the code match the rule?**

Compare against the quoted source sentence. `no` if the code demands something the
rule does not, accepts something the rule forbids, or narrows the rule to a special
case. A threshold that differs from the one in the text is a `no`, and so is an
exception the rule grants that the code ignores.

**C2. Does the precondition pick the right things?**

A rule about commits should return commits, and a rule about added functions should
return added functions. `no` if it returns nothing for work the rule plainly covers,
which scores the rule as inapplicable rather than violated, or if it returns things
the rule does not cover, which marks unrelated work as non-compliant. Where the row
has a trigger condition, check that the precondition is testing that condition.

**C3. Does it actually behave correctly?**

Run the check against the cases in its row. The satisfying case should come back
satisfied, the violating case violated, and the inapplicable case inapplicable. `no`
if any of them is wrong, if the check crashes, or if you can think of a realistic
patch it would obviously get wrong. Note that patch if so.

Some rows are missing a case. Leave that column blank and say so in the notes.

**C4. Is the check exact or approximate?**

Answer `exact` or `approximate`, the same two labels the paper uses.

An exact check decides its rule, as counting the characters in a commit summary does.
An approximate one stands in for a reading the documentation does not fully specify,
as looking for words ending in -ed does for a rule about the past tense. Approximate
is acceptable and most checks are, so what you are judging is whether the label
already on the row matches what the code actually does.

## The verdict

One word, `accept` or `reject`.

`accept` says the code is a correct encoding of its rule, meaning C1 to C3 are all
`yes` and C4 matches the label already on the row.

`reject` says it is not. Name the failing question in the next column and say what is
wrong in the notes.

There is no middle label. If you cannot settle an item, reject it and say why, because
`accept` is a statement that the check is correct and a rater who is unsure is not in a
position to make it. Whether a rejected check can be fixed with a small edit or has to
be removed is decided after adjudication, not by you while labelling.

## Notes

One line. What is wrong with the check, the patch you thought of, a missing case, or a
problem with the rule rather than the code.

Do not write a note that would fit any other item. A note that generic means the
verdict came from pattern rather than reading.

## Do not edit the checks

Leave the descriptive columns alone and do not fix a check while reviewing it. A
repaired check cannot be labelled by the other rater. Note the ID and move on.

## Afterwards

Both sheets are compared item by item. Every item where the verdicts differ, or where
C1, C2 or C3 differ, goes to a joint session.

\end{Verbatim}
\end{tcolorbox}
\subsection{Construction Prompts}
\label{app:prompts}

Each repository was handed to an agent with the launcher prompt below, with the organisation, repository, and slug filled in. The launcher directs the agent to the workflow document, which carries one instruction block per stage, together with the two rubrics (Appendix~\ref{app:rubrics}), the source-discovery script, and a format-standard workbook. The launcher and the three stage blocks are reproduced below. So that the prompts read consistently with the rest of the paper, field names and label values have been renamed to the paper's terminology; for example, the working field names for scope and Evidence Type are replaced by those terms, and the requirement levels appear as \textit{must} and \textit{should}. Non-ASCII punctuation has also been replaced with ASCII equivalents. The prompts are otherwise verbatim, and the unedited originals, together with the complete workflow document, are released with the dataset. The reviewer checkpoints that follow each block are summarised in Table~\ref{tab:checkpoints}, and placeholders in angle brackets were filled per repository. The last block (Appendix~\ref{app:prompt-checkers}) is the guide the agent followed when it wrote the checker functions of Appendix~\ref{app:checkers}.

\subsubsection{Launcher prompt}
\label{app:prompt-launcher}
\begin{tcolorbox}[breakable, colback=gray!4, colframe=gray!45, boxrule=0.4pt, left=4pt, right=4pt]
\begin{Verbatim}[fontsize=\scriptsize, breaklines=true, breakanywhere=true]
You are extracting and classifying contribution rules for exactly ONE repository:
**`<org>/<repo>`**, slug **`<slug>`**.

## Where the work happens

The repository root is `<repo_root>`. Anchor every path there. Split long-running
jobs into steps and keep scratch files outside the repository root.

Use web retrieval (fetch and search) to read the documentation pages.

## What to read, in this order

1. `rule-extraction/rule-extraction-workflow-v2.md` -- the method, end to end. Follow it.
2. `rule-extraction/care-rubric.md` -- Rubric A, applied at Part 3 step 2.
3. `rule-extraction/strength-rubric.md` -- Rubric B, applied at Part 3 step 4.
4. `rule-extraction/seed_sources.py` -- read the docstring and `PATTERNS`.
5. `rules/sympy/sympy_rules.xlsx` -- the format standard. Match its shape.
6. `rules/sympy/repo.conf` -- the per-repo config you will write.

Do not start Part 1 until you have read all six.

## Constraints for this run

- **No git commands.** Sibling agents share this working tree and it is already dirty.
  Write files only; the orchestrator commits.
- **Touch nothing outside `rules/<slug>/`.** Not another repo's directory, not
  `RULE_MODULES`, not an existing corpus. You have no mandate to fix other sheets.
- **Do not re-decide anything the workflow doc settles.** Where your judgment differs,
  say so in the run log and follow the doc.
- The conflict tiebreaker is fixed centrally: **the stricter statement governs, the weaker
  one is N3 with the conflict named in `Notes`.** Not yours to re-open.

## Done means

Both gates pass, per Part 4:

```
python tools/corpus.py --repo <slug>
python tools/audit_rules.py --repo <slug>
```

`tools/audit_rules.py` already exists -- use it, do not rewrite it.

## Report back

Under 350 words: the audit pass/fail table, row count, in scope/out of scope split, must/should
split, the four N-route counts, and one paragraph on what this repo's prose did that the
others' did not. That paragraph is the actual research output; the workbook is its
evidence. Flag anything in the workflow doc that was wrong, ambiguous, or missing.
\end{Verbatim}
\end{tcolorbox}

\subsubsection{Stage 1 prompt: source manifest}
\label{app:prompt-manifest}
\begin{tcolorbox}[breakable, colback=gray!4, colframe=gray!45, boxrule=0.4pt, left=4pt, right=4pt]
\begin{Verbatim}[fontsize=\scriptsize, breaklines=true, breakanywhere=true]
You are building a source manifest for `<REPO>`, ahead of extracting contribution rules.

Docs root: `<PINNED DOCS ROOT>`

**Fetch every page you assess.** Do not judge from nav titles, and do not use prior knowledge of this repo. If a fetch fails, mark the row `FETCH FAILED` and continue. Never assert the contents of a file you did not fetch.

Record the version string the docs serve. Any page on a different version is `VERSION MISMATCH`, not a row. Report it and stop.

### Scope test

Apply literally to every page and section:

> Could a rule here be violated by the diff, the commit message, or the files the agent writes?

In scope: coding style, testing, documentation and docstrings, workflow, commit and PR conventions, deprecation, automated-contribution or AI policy.

### Traversal

Walk the navigation, then take **one hop** out of in-scope sections. Repo-internal targets only. Do not recurse further.

| Target | Action |
| --- | --- |
| Prose stating obligations | fetch, assess, add to manifest |
| Config file (`.pre-commit-config.yaml`, `pyproject.toml`) | context-only, no rows |
| Source file (`conftest.py`, `api_reference.py`) | do not fetch, name it in Notes |
| External standard (PEP8, numpydoc, PEP440) | one pointer rule, stop |
| Sibling repo under the same org (`astropy/astropy-project`, `pallets/website`) | in scope -- fetch, assess, name the source repo in the row |
| Agent-instruction file (`CLAUDE.md`, `AGENTS.md`, `.github/copilot-instructions.md`, an AI policy) | in scope wherever it exists, even though it sits outside the docs navigation |

### Roles

Every source gets `rules`, `context-only`, or `excluded`.

`context-only` is a real category, not a workaround. A changelog legend defining what each fragment type means produces zero rules but is what makes fragment-type selection decidable rather than a judgment call. A pre-commit config produces zero rules but decides the auto-fix reading. Marking these `excluded` loses the fact you consulted them.

### Exclusion codes

| Code | Means |
| --- | --- |
| `X-INSTALL` | install, build, environment setup |
| `X-GOV` | governance, community process |
| `X-MAINT` | maintainer-only: releasing, merging, version bumps, website |
| `X-TRIAGE` | triaging issues, reviewing other people's PRs |
| `X-ISSUE` | constrains an issue body or bug report, not a contribution |
| `X-NARRATIVE` | rationale, history, tips, no obligation |

Keep `X-ISSUE` separate from `X-NARRATIVE`. Bug report requirements are real obligations the rig cannot reach. Folding them into narrative claims they were never rules at all.

### Mixed pages

A page is not all-or-nothing. Where one page carries both in-scope and out-of-scope material, mark it `Partial` and name the section headings on each side.

This is the common case, not the exception. Repos routinely put contributor workflow and maintainer or triage guidance under one page.

### Output

`Page | URL | Role | Include (Full/Partial/Exclude) | In-scope sections | Code | Reason`

Then stop. Do not extract anything. I will confirm the manifest first.
\end{Verbatim}
\end{tcolorbox}

\subsubsection{Stage 2 prompt: extraction}
\label{app:prompt-extraction}

Supplied with the confirmed manifest and the bundle of raw off-navigation sources.

\begin{tcolorbox}[breakable, colback=gray!4, colframe=gray!45, boxrule=0.4pt, left=4pt, right=4pt]
\begin{Verbatim}[fontsize=\scriptsize, breaklines=true, breakanywhere=true]
Extract atomic contribution rules from the confirmed manifest. Work **one source at a time**, largest first.

The attached file contains raw source for off-nav files, HTML comments intact. Extract those from the attachment, not from any rendered page.

Fetch each doc page in full and read it semantically. No keyword search. No prior knowledge of this repo. Do not reconstruct verbatim text from memory. If earlier pages have dropped out of your context, re-fetch them.

Restate the manifest row before each source: page, URL, in-scope sections. On a `Partial` page extract only from the in-scope sections. Read the rest for section context where it governs an in-scope rule, but produce no rows from it.

### Rule definition

A repo-specific requirement, prohibition, or recommendation about **creating, validating, or submitting a contribution**. Not description, rationale, history, or advice about reviewing someone else's work.

### Atomicity

One condition plus one expected behaviour per row. Split bundled obligations. Keep conditions attached to their rule: do not strip "first-time contributors only" off the rule it governs.

### Extraction columns

These are working columns for Part 2 only. Part 3 folds them into the final eleven.

| Column | Content |
| --- | --- |
| ID | `<PREFIX>-C001`, sequential across the whole run |
| Source | **full URL first**, then section: `<URL> \| <Section>`. Add ` \| <Subsection>` only where the page needs it to be unambiguous. |
| Original text | verbatim. Never paraphrase this column. |
| Atomic rule | one imperative sentence, your words |
| Modal-verb level | naive modal reading of the source only: `must` / `should` / `prohibited`. Hedged and preference modals ("should", "try to", "prefer") read as `should`; this column shares the final vocabulary so Part 5 can diff the two. |
| Applies to | `code`, `test`, `docs`, `commit`, `PR`, `workflow`, `CI`. Comma-join where a rule genuinely spans two. `PR` and `CI` are uppercase. |
| Category | Documentation and docstrings / Tests and test style / Specialized changes / PR and release metadata / Git and commit conventions / AI-assisted contribution policy / Code and quality / Language and framework style |
| Section context | surrounding text that grants an exception, states a consequence, names enforcement tooling, or offers an alternative. Quote it. `none` if the section grants nothing. |
| Notes | trigger conditions, cross-references |

**Modal-verb level is a working column, not an output column.** It is the naive modal reading and it is what Part 3 measures its own answer against. It never reaches the sheet as a column; Part 3 records it inside `Notes`. `prohibited` is legal here and only here.

**`Category` is a closed list.** Do not coin a repo-specific label. Repo-specific style rules (a framework's template, model, or view conventions; a language's import ordering) all go to `Language and framework style`, which is deliberately broad. A repo with no such rules leaves the bucket empty.

### Section context is not optional

Requirement level turns on it. Django's C047 (72 chars, `should`) and SymPy's C023 (71 chars, `must`) are near-identical clauses. The split sits outside the quoted text, in whether the surrounding docs sanction deviation. The clause alone cannot decide.

### Triggers must be explicit

Any rule conditional on the contribution touching a subsystem gets its condition written into `Notes` in plain form:

> Applies only if the PR adds or modifies a Display class.
> Applies only if the PR contains Cython.

Classification cannot separate N4 from `PassType: never fires` without this.

### API-contract pages are a different shape

Interface specs are not instruction prose. "The Display class exposes `from_estimator`" is an API inventory, which under the rule definition is description, not obligation. The rule is the conformance requirement: *if you implement a Display, it must expose these.*

Extract the conformance obligation once, at the right level. Do not emit one row per protocol member, attribute, or method signature. If you are producing a row that restates a member of an interface, you are cataloguing an API rather than extracting rules.

### Zero-rule sources

If an in-scope section yields no rules, say so explicitly with the reason. Do not silently omit it. A source that produced nothing and a source you forgot look identical in the sheet.

### Per source

Report the row count, then stop and wait.

### Closing report

When every source is done:

1. Final row count and ID range
2. Contiguity: no gaps, no repeats in the ID sequence
3. Exact-duplicate `Original text` values across all sources
4. Rows per source, including any that yielded zero
5. Modal-verb level, `Applies to`, `Category` distributions

Then stop.
\end{Verbatim}
\end{tcolorbox}

\subsubsection{Stage 3 prompt: classification}
\label{app:prompt-classification}

The block embeds the two rubrics of Appendix~\ref{app:rubrics}. We omit them here and reproduce the rest of the block.

\begin{tcolorbox}[breakable, colback=gray!4, colframe=gray!45, boxrule=0.4pt, left=4pt, right=4pt]
\begin{Verbatim}[fontsize=\scriptsize, breaklines=true, breakanywhere=true]
You are classifying already-extracted rules. Input rows carry ID, Source, Original text, Atomic rule, modal-verb level, Applies to, Category, Section context, Notes. Nothing else.

Work in batches of 25, **across the whole sheet, not per page.** Duplicate detection needs to see rules from different pages side by side. Run per page and you never catch that an 80 percent coverage rule duplicates a 90 percent one.

### Order

Do not skip ahead.

1. `Evidence Type`
2. `In scope` via Rubric A
3. `PassType` on in-scope rows
4. `Requirement level` via Rubric B, in-scope rows only
5. `DecidedBy`
6. Reasoning
7. Auto-fix reading

\end{Verbatim}
\end{tcolorbox}

\textit{[The block embeds Rubric A (scope) and Rubric B (requirement level) here, as given in Appendix~\ref{app:rubrics}.]}

\begin{tcolorbox}[breakable, colback=gray!4, colframe=gray!45, boxrule=0.4pt, left=4pt, right=4pt]
\begin{Verbatim}[fontsize=\scriptsize, breaklines=true, breakanywhere=true]
### Auto-fix reading

`Yes, auto-fixed` / `No, check only` / `Partial risk`. Only where a formatter or hook can silently repair the violation before the patch exists. Otherwise omit it.

Decide from the repo's actual hook config in the attached raw sources, not from what a tool is generally capable of. A linter invoked without its fix flag is `No, check only`.

This is no longer a column. Where it applies, it goes into `Notes` as a trailing clause: `Auto-fix risk: Yes, auto-fixed -- black rewrites the file on commit.` Where it does not apply, write nothing.

### Composing the `Notes` cell

`Notes` carries everything the eighteen-column draft used to spread across `Section context`, `PassType`, `DecidedBy`, `Evidence & Reasoning`, `Auto-fix risk`, and `Auto-fix note`. Fixed order, pipe-separated, so it stays machine-splittable:

**In-scope rows**

```
DecidedBy: <M1-M6|D1-D4> | PassType: <checked|never fires|always fails> | Modal-verb level: <must|should|prohibited> | Quote: "<under 15 words, verbatim>" | Context: <what the section granted or withheld> | Conclusion: <one clause>
```

**Out-of-scope rows**

```
Modal-verb level: <must|should|prohibited> | Quote: "<under 15 words, verbatim>" | Context: <what the section granted or withheld> | Conclusion: <one clause>
```

Append these where they apply, in this order:

- `| Trigger: <plain-form condition>` on any N4 or `never fires` row
- `| Auto-fix risk: <reading> -- <one line>`
- `| Conflict: <ID> -- <stricter statement governs>` on an N3 conflict row
- `| low_confidence: <why>` wherever you guessed

**`Modal-verb level` is not optional.** It is the only surviving record of the naive modal reading, and Part 5's promotion-direction check is computed from it. Dropping it costs the research signal the two stages exist to produce. Record it on every row, in scope and out of scope alike, including where it equals the final `Requirement level`.

**Every row gets distinct reasoning.** If two rows would read identically, either they are duplicates (N3) or you have defaulted to boilerplate. Boilerplate made an earlier sheet unauditable and is a failure of this task.

### Output columns

Eleven, in this order, matching `sympy_rules.xlsx` exactly:

```
ID | Source | Original text | Atomic rule | Applies to | Category | Requirement level | Evidence Type | In scope | Out-of-scope reason | Notes
```

| Column | Content |
| --- | --- |
| ID | `<PREFIX>-C001`, carried from extraction |
| Source | `<URL> \| <Section>`, URL first |
| Original text | verbatim, carried from extraction |
| Atomic rule | carried from extraction |
| Applies to | `code`, `test`, `docs`, `commit`, `PR`, `workflow`, `CI` |
| Category | one of the eight |
| Requirement level | `must` / `should`. **Blank on every out-of-scope row.** |
| Evidence Type | output / differential / trajectory / judgment |
| In scope | `TRUE` / `FALSE` |
| Out-of-scope reason | `N1 not observable` / `N2 subjective` / `N3 duplicate` / `N4 external condition`. **Blank on every in-scope row.** |
| Notes | composed as above |

Four columns from the earlier draft are gone as columns and live inside `Notes`: `Section context`, `PassType`, `DecidedBy`, `Auto-fix risk` / `Auto-fix note`. `Final Requirement level` is now just `Requirement level`; the naive reading moves to `Modal-verb level:` in `Notes`. The out-of-scope reason column carries the N-code in the cell text.

### Per batch

Counts by Evidence Type, In scope, Out-of-scope reason, Requirement level, DecidedBy, PassType, plus the number of rows where `Modal-verb level != Requirement level`.
\end{Verbatim}
\end{tcolorbox}

\subsubsection{Checker authoring guide}
\label{app:prompt-checkers}

Checker functions are authored from a standing guide rather than from a per-repository prompt.
The agent for a repository is pointed at the guide and follows it end to end: it states the
two-function contract, how the condition and behaviour halves of a policy statement map onto the
precondition and the post-condition, how the outcomes on individual targets combine into one outcome for the policy,
the three synthetic inputs every function is tested on, and what to record when a policy cannot be coded at all. The
guide names no project, and the same text is used for all twelve. It is released with the
benchmark; the three clauses that decide most authoring questions are reproduced below.

\begin{tcolorbox}[breakable, colback=gray!4, colframe=gray!45, boxrule=0.4pt, left=4pt, right=4pt]
\begin{Verbatim}[fontsize=\scriptsize, breaklines=true, breakanywhere=true]
[Contract]
precondition(bundle) -> list[Target]
    Answers *does this policy apply to what the agent did?* It selects and judges nothing.
post_condition(target) -> Satisfied | Violated | Undetermined
    Answers *given that it applies, did the agent get it right?* It runs once per target,
    reads only that target's payload, and never re-selects. If it wants different material,
    the precondition is wrong.
    Undetermined(reason) -- the policy applies but the artifacts cannot answer it.
    Never a violation, never a silent pass.

[Selection]
precondition selects on the ANTECEDENT -- the situation that invokes the policy -- not on
the artefact the policy demands. "Do not force-push" is: precondition every `git push`;
post-condition no force flag. Selecting force-pushes would only ever find violations and
could never record a compliant push.

[Tests]
Three per policy: satisfied, violated, and an input where the precondition finds nothing.
Be honest about the limit: these tests PIN your reading of the sentence; they do not
VALIDATE it. Both would pass on a predicate that confidently grades the wrong thing.
\end{Verbatim}
\end{tcolorbox}

\subsection{Agent Prompts}
\label{app:agent-prompts}

Both scaffolds run with their stock SWE-bench configuration. We keep each scaffold's system prompt and replace its SWE-bench task prompt with a modified copy. The copy makes three changes. It widens the task to include tests and documentation, adds a Compliance section that carries the condition treatment, and adds a Submission section that asks for a git commit and a pull-request description. The Compliance paragraphs are identical in the two scaffolds. In the task prompts below, highlighted lines are added or reworded relative to the stock prompt. Each run receives one of the two Compliance paragraphs, and we show both with a label. Placeholders in angle brackets are filled per instance, and non-ASCII characters are replaced with ASCII equivalents. The prompts are otherwise verbatim.

\subsubsection{mini-SWE-agent}
\label{app:prompt-msa}

\textbf{System prompt.} Unchanged from the stock SWE-bench configuration.

\begin{tcolorbox}[breakable, colback=gray!4, colframe=gray!45, boxrule=0.4pt, left=4pt, right=4pt]
\begin{Verbatim}[fontsize=\scriptsize, breaklines=true, breakanywhere=true]
You are a helpful assistant that can interact with a computer shell to solve programming tasks.
\end{Verbatim}
\end{tcolorbox}

\textbf{Task prompt.} The stock prompt restricts the agent to non-test files, forbids changes to tests and configuration, tells the agent it may install missing tools, and asks it to submit an uncommitted \texttt{patch.txt}. The modified prompt removes these instructions and adds the highlighted lines. We also remove the stock step and cost limits, and we exempt the rules file from the scaffold's 10,000-character output truncation, which otherwise drops the middle of the file.

\begin{tcolorbox}[breakable, colback=gray!4, colframe=gray!45, boxrule=0.4pt, left=4pt, right=4pt]
\begin{Verbatim}[fontsize=\scriptsize, breaklines=true, breakanywhere=true, highlightlines={12-13,23-29,36-38,75-101,105-115,118,122-130}, highlightcolor=promptchange]
<pr_description>
Consider the following PR description:
<problem statement>
</pr_description>

<instructions># Task Instructions

## Overview

You're a software engineer interacting continuously with a computer by submitting commands.
You'll be helping implement necessary changes to meet requirements in the PR description.
Your task is to make changes in the current directory that fix the issue described in the PR
description in a way that is general and consistent with the codebase.
<IMPORTANT>This is an interactive process where you will think and issue AT LEAST ONE command, see the result, then think and issue your next command(s).</important>

For each response:

1. Include a THOUGHT section explaining your reasoning and what you're trying to accomplish
2. Provide one or more bash tool calls to execute

## Important Boundaries

- MODIFY: source files, test files, and documentation in /testbed (this is the working
  directory for all your subsequent commands). If your change adds or alters behaviour,
  add or update tests for it in the repository's own test files, following this
  repository's testing conventions.
- DO NOT delete, weaken, or skip existing tests in order to make them pass.
- DO NOT MODIFY: packaging and CI configuration (pyproject.toml, setup.py, setup.cfg,
  tox.ini, .github/) unless the task genuinely requires it.

## Recommended Workflow

1. Analyze the codebase by finding and reading relevant files
2. Create a script to reproduce the issue
3. Edit the source code to resolve the issue
4. Add or update tests covering the change
5. Verify your fix works by running your script and the relevant tests again
6. Test edge cases to ensure your fix is robust

## Command Execution Rules

You are operating in an environment where

1. You issue at least one command
2. The system executes the command(s) in a subshell
3. You see the result(s)
4. You write your next command(s)

Each response should include:

1. **Reasoning text** where you explain your analysis and plan
2. At least one tool call with your command

**CRITICAL REQUIREMENTS:**

- Your response SHOULD include reasoning text explaining what you're doing
- Your response MUST include AT LEAST ONE bash tool call. You can make MULTIPLE tool calls in a single response when the commands are independent (e.g., searching multiple files, reading different parts of the codebase).
- Directory or environment variable changes are not persistent. Every action is executed in a new subshell.
- However, you can prefix any action with `MY_ENV_VAR=MY_VALUE cd /path/to/working/dir && ...` or write/load environment variables from files

Example of a CORRECT response:
<example_response>
I need to understand the Builder-related code. Let me find relevant files and check the project structure.

[Makes multiple bash tool calls: {"command": "ls -la"}, {"command": "find src -name '*.java' | grep -i builder"}, {"command": "cat README.md | head -50"}]
</example_response>

## Environment Details

- You have a full Linux shell environment
- Always use non-interactive flags (-y, -f) for commands
- Avoid interactive tools like vi, nano, or any that require user input
- You can use bash commands or invoke any tool that is available in the environment
- You can also create new tools or scripts to help you with the task
- `git` is configured with a working identity (`user.name`/`user.email`), so a real commit is
  possible. There is no `gh` CLI and no remote to push to. You will not push, and you will not
  open a real pull request -- your PR submission is simulated as text (see Submission below),
  not a real GitHub object.
- This environment **has outbound network access**, so you may fetch pages from the web
  (for example with `curl` or `python -c "import urllib.request; ..."`).

## Compliance
[Native condition]
This repository publishes contribution guidelines online, at:

    <documentation URL>

**Fetch and read those guidelines before you start any work**, and follow them. Follow the
links from that index page to the sections relevant to your change.

[Consolidated condition]
This repository's contribution guidelines have been extracted for you and placed in the
container at `/rules/CONTRIBUTING_RULES.md`.

**Read `/rules/CONTRIBUTING_RULES.md` before you start any work**, and follow it. That file
is this project's contribution guidelines, extracted in full.

In addition:

- Create an actual Git commit for your work.
- Write a pull-request description for it (see Submission below).

## Submission

When you are done, submit your work as a real git commit plus a written pull-request
description. There is no real GitHub PR in this environment -- the pull request is text
only. Follow these steps, as SEPARATE commands:

1. Commit the changes that belong in the contribution, with `git commit` (a real
   commit, not just a diff). Decide for yourself what belongs in it and what does
   not -- scratch files you created while investigating are yours to keep or clean
   up, and staging is your decision, not ours.
2. In the same response as step 3, write your pull-request description in your THOUGHT
   text, starting with the exact line `PR SUBMISSION:`.
3. Submit, as its own separate command -- run this exact command, unmodified:

```bash
echo COMPLETE_TASK_AND_SUBMIT_FINAL_OUTPUT && bash /opt/collect.sh
```

<CRITICAL>
- Step 1 (commit) and step 3 (submit) must each be their own separate tool call -- do not
  chain them together.
- The submit command in step 3 is the one exception to that rule: run it exactly as shown,
  including `&& bash /opt/collect.sh`. `/opt/collect.sh` is what produces the patch and the
  commit metadata that your work is graded on. Do not edit it, do not replace it with
  `cat`, and do not omit it.
- You do not need to build a patch file yourself. `/opt/collect.sh` diffs the repository
  for you, including any files you created.
- You CANNOT continue working (reading, editing, testing) on this task after submitting.
</CRITICAL>
</instructions>
\end{Verbatim}
\end{tcolorbox}

\subsubsection{OpenHands}
\label{app:prompt-openhands}

\textbf{System prompt.} We use the SDK's stock system prompt at the pinned commit (\texttt{software-agent-sdk@43376f1}), which we do not reproduce in full. Upstream disables the browser for SWE-bench. We enable it so that the agent can fetch policy pages under \textit{Native}. This adds the browser toolset and the \texttt{<BROWSER\_TOOLS>} block below (highlighted), which is part of the stock prompt and appears only when the browser is enabled. We also show the stock \texttt{<EXTERNAL\_SERVICES>} block, which already asks the agent to disclose AI assistance in content it posts. For GPT-5.6 Luna and Gemini 3.7 Flash, the SDK appends a short model-specific block. The SDK also appends the current date and the list of available skills. The full prompt and these additions are released with the trajectories.

\begin{tcolorbox}[breakable, colback=gray!4, colframe=gray!45, boxrule=0.4pt, left=4pt, right=4pt]
\begin{Verbatim}[fontsize=\scriptsize, breaklines=true, breakanywhere=true, highlightlines={1-9}, highlightcolor=promptchange]
<BROWSER_TOOLS>
You have a browser for navigating pages and interacting with web UIs.
* Try curl/wget/fetch first. Use the browser only when simpler tools fail or the page requires JS/interaction.
* ALWAYS call `browser_get_state` before EVERY `browser_click` or `browser_type` -- indices change after each action. Flow: navigate -> get_state -> interact -> get_state -> get_content.
* Max 10 browser actions per sub-task. If stuck, switch approach entirely.
* If 20+ total steps without converging, stop exploring and commit to your best answer.
* On 403/CAPTCHA/login wall: try one alternative, then abandon the browser.
* Do NOT submit forms or create accounts unless explicitly asked.
</BROWSER_TOOLS>

<EXTERNAL_SERVICES>
* When interacting with external services like GitHub, GitLab, or Bitbucket, use their respective APIs instead of browser-based interactions whenever possible.
* Only resort to browser-based interactions with these services if specifically requested by the user or if the required operation cannot be performed via API.
* **AI disclosure**: When posting messages, comments, issues, or any content to external services that will be read by humans (e.g., Slack messages, GitHub/GitLab comments, PR/MR descriptions, Discord messages, Linear/Jira issues, Notion pages, emails, etc.), always include a brief note indicating the content was generated by an AI agent on behalf of the user. For example, you could add a line like: _"This [message/comment/issue/PR] was created by an AI agent (OpenHands) on behalf of [user]."_ This applies to any communication channel -- whether through dedicated tools, MCP integrations, or direct API calls.
</EXTERNAL_SERVICES>
\end{Verbatim}
\end{tcolorbox}

\textbf{Task prompt.} The stock prompt tells the agent that test changes are already handled and asks for minimal changes to non-test files. The modified prompt replaces these two sentences and adds the highlighted lines.

\begin{tcolorbox}[breakable, colback=gray!4, colframe=gray!45, boxrule=0.4pt, left=4pt, right=4pt]
\begin{Verbatim}[fontsize=\scriptsize, breaklines=true, breakanywhere=true, highlightlines={8-16,71-124}, highlightcolor=promptchange]
I have access to a python code repository in the directory <repo path> . You can explore and modify files using the available tools. Consider the following issue description:

<issue_description>
<problem statement>
</issue_description>

Can you help me implement the necessary changes to the repository so that the requirements specified in the <issue_description> are met?
The development Python environment is already set up for you (i.e., all dependencies already installed), so you don't need to install other packages.
Your task is to make changes in the <repo path> directory that fix the issue described in the <issue_description> in a way that is general and consistent with the codebase.

- MODIFY: source files, test files, and documentation in <repo path>. If your
  change adds or alters behaviour, add or update tests for it in the repository's own test
  files, following this repository's testing conventions.
- DO NOT delete, weaken, or skip existing tests in order to make them pass.
- DO NOT MODIFY: packaging and CI configuration (pyproject.toml, setup.py, setup.cfg,
  tox.ini, .github/) unless the task genuinely requires it.

Follow these phases to resolve the issue:

Phase 1. READING: read the problem and reword it in clearer terms
   1.1 If there are code or config snippets. Express in words any best practices or conventions in them.
   1.2 Hightlight message errors, method names, variables, file names, stack traces, and technical details.
   1.3 Explain the problem in clear terms.
   1.4 Enumerate the steps to reproduce the problem.
   1.5 Hightlight any best practices to take into account when testing and fixing the issue

Phase 2. RUNNING: install and run the tests on the repository
   2.1 Activate the environment by running  
       source /opt/miniconda3/etc/profile.d/conda.sh ; conda activate testbed
   2.2 Follow the readme
   2.3 Install the environment and anything needed
   2.4 Iterate and figure out how to run the tests

Phase 3. EXPLORATION: find the files that are related to the problem and possible solutions
   3.1 Use `grep` to search for relevant methods, classes, keywords and error messages.
   3.2 Identify all files related to the problem statement.
   3.3 Propose the methods and files to fix the issue and explain why.
   3.4 From the possible file locations, select the most likely location to fix the issue.

Phase 4. TEST CREATION: before implementing any fix, create a script to reproduce and verify the issue.
   4.1 Look at existing test files in the repository to understand the test format/structure.
   4.2 Create a minimal reproduction script that reproduces the located issue.
   4.3 Run the reproduction script to confirm you are reproducing the issue.
   4.4 Adjust the reproduction script as necessary.

Phase 5. FIX ANALYSIS: state clearly the problem and how to fix it
   5.1 State clearly what the problem is.
   5.2 State clearly where the problem is located.
   5.3 State clearly how the test reproduces the issue.
   5.4 State clearly the best practices to take into account in the fix.
   5.5 State clearly how to fix the problem.

Phase 6. FIX IMPLEMENTATION: Edit the source code to implement your chosen solution.
   6.1 Make minimal, focused changes to fix the issue.

Phase 7. VERIFICATION: Test your implementation thoroughly.
   7.1 Run your reproduction script to verify the fix works.
   7.2 Add edge cases to your test script to ensure comprehensive coverage.
   7.3 Run existing tests related to the modified code to ensure you haven't broken anything.

8. FINAL REVIEW: Carefully re-read the problem description and compare your changes with the base commit <base commit>.
   8.1 Ensure you've fully addressed all requirements.
   8.2 Run any tests in the repository related to:
     8.2.1 The issue you are fixing
     8.2.2 The files you modified
     8.2.3 The functions you changed
   8.3 If any tests fail, revise your implementation until all tests pass

Be thorough in your exploration, testing, and reasoning. It's fine if your thinking process is lengthy - quality and completeness are more important than brevity.

## Compliance
[Native condition]
This repository publishes contribution guidelines online, at:

    <documentation URL>

**Fetch and read those guidelines before you start any work**, and follow them. Follow the
links from that index page to the sections relevant to your change.

[Consolidated condition]
This repository's contribution guidelines have been extracted for you and placed in the
container at `/workspace/.compliance/rules/CONTRIBUTING_RULES.md`.

**Read `/workspace/.compliance/rules/CONTRIBUTING_RULES.md` before you start any work**, and follow it. That file
is this project's contribution guidelines, extracted in full.

In addition:

- Create an actual Git commit for your work.
- Write a pull-request description for it (see Submission below).

## Submission

When you are done, submit your work as a real git commit plus a written pull-request
description. There is no real GitHub PR in this environment -- the pull request is text
only. `git` is configured with a working identity, so a real commit is possible; there is
no `gh` CLI and no remote to push to, so you will not push and will not open a real pull
request.

Follow these steps, as SEPARATE tool calls:

1. Commit the changes that belong in the contribution, with `git commit` (a real
   commit, not just a diff). Decide for yourself what belongs in it and what does
   not -- scratch files you created while investigating are yours to keep or clean
   up, and staging is your decision, not ours.
2. Run this exact command, unmodified, in the terminal:

```bash
bash /workspace/.compliance/collect
```

3. Finish, putting your pull-request description in the final message, starting with the
   exact line `PR SUBMISSION:`.

<CRITICAL>
- Steps 1, 2 and 3 must each be their own separate tool call -- do not chain them.
- Run `/workspace/.compliance/collect` exactly as shown. It is what produces the patch and the commit
  metadata your work is graded on. Do not edit it, do not replace it with `git diff`, and
  do not omit it.
- You do not need to build a patch file yourself. That script diffs the repository
  for you, including any files you created.
- Run it AFTER your commit and BEFORE you finish, or your work is not
  recorded.
</CRITICAL>
\end{Verbatim}
\end{tcolorbox}

\end{document}